\documentclass[aps,prl,twocolumn,titlepage,superscriptaddress,nofootinbib]{revtex4-1}

\usepackage{graphicx}     
\usepackage{amssymb}      
\usepackage{amsmath}     
\usepackage{latexsym}     
\usepackage{cancel}       
\usepackage[normalem]{ulem} 
\usepackage{url}          
\usepackage{soul}          
\usepackage{verbatim}     
\usepackage{multirow}     
\usepackage{mathrsfs}     
\usepackage{float}         
\usepackage[dvipsnames]{xcolor} 
\usepackage{mathtools}    
\usepackage{slashed}       
\usepackage{physics}      
\usepackage{epstopdf}     
\usepackage{subfigure}   
\usepackage{bbold}        
\usepackage{wasysym}      
\usepackage{feynmp}      

\usepackage[colorlinks=true, citecolor=blue, linkcolor=blue, urlcolor=blue]{hyperref}
\usepackage{orcidlink}
\usepackage{etoolbox}

\newcommand{\beq}{\begin{eqnarray}}
\newcommand{\eeq}{\end{eqnarray}}

\makeatletter
\newcommand{\myBig}{\bBigg@{1.75}}
\makeatother

\begin{document}
\title{Universal fingerprint of topological defect cores}

\author{Zi-Qiang Zhao\orcidlink{0009-0009-7859-3655}}
\affiliation{Liaoning Key Laboratory of Cosmology and Astrophysics, College of Sciences, Northeastern University, Shenyang 110819, China}
\author{Nayun Jia\orcidlink{0009-0009-6136-2682}}
\affiliation{Liaoning Key Laboratory of Cosmology and Astrophysics, College of Sciences, Northeastern University, Shenyang 110819, China}
\author{Zhang-Yu Nie\orcidlink{0000-0001-7064-247X}}\email{niezy@kust.edu.cn}
\affiliation{Center for Gravitation and Astrophysics, Kunming University of Science and Technology, Kunming 650500, China}
\author{Hua-Bi Zeng\orcidlink{0000-0001-7409-0537}}\email{zenghuabi@hainanu.edu.cn}
\affiliation{Center for Theoretical Physics , Hainan University, Haikou 570228, China}
\author{Jing-Fei Zhang\orcidlink{0000-0002-3512-2804}}
\affiliation{Liaoning Key Laboratory of Cosmology and Astrophysics, College of Sciences, Northeastern University, Shenyang 110819, China}
\author{Xin Zhang\orcidlink{0000-0002-6029-1933}}\email{zhangxin@neu.edu.cn}
\affiliation{Liaoning Key Laboratory of Cosmology and Astrophysics, College of Sciences, Northeastern University, Shenyang 110819, China}
\affiliation{MOE Key Laboratory of Data Analytics and Optimization for Smart Industry, Northeastern University, Shenyang 110819, China}
\affiliation{National Frontiers Science Center for Industrial Intelligence and Systems Optimization, Northeastern University, Shenyang 110819, China}

\begin{abstract}
Topological defects are ubiquitous in physics, arising from condensed matter physics to the early universe. Although there exist many universal scaling laws for correlations between topological defects, such as Porod scaling, the fingerprint of topological defect cores has remained largely unexplored. Here, we discover a universal scaling law in the region $k>1/\xi$, where $\xi$ is the healing length of the topological defects, taking the scaling of the form factor $S_f \propto k^{-(d+p+2)}$, where $d$ is the spatial dimension and $p$ is the defect codimension. We analytically prove that this exponent originates from a universal V-shaped cusp at the defect core and is independent of the underlying system and dynamics. Numerical simulations verify this scaling law in four typical frameworks: the time-dependent Ginzburg-Landau and Gross-Pitaevskii equations in the weak-coupling regime, the gauge/gravity duality model in the strong-coupling regime, and the Klein-Gordon equation in the Friedmann-Robertson-Walker background in cosmology. Our work provides a new probe for studying topological defects in systems ranging from superconductors to cosmological phase transitions.
\end{abstract}
\maketitle

\section{Introduction}
\begin{figure*}
    \centering
    \subfigure[]{\label{kink}\includegraphics[width=0.55\columnwidth]{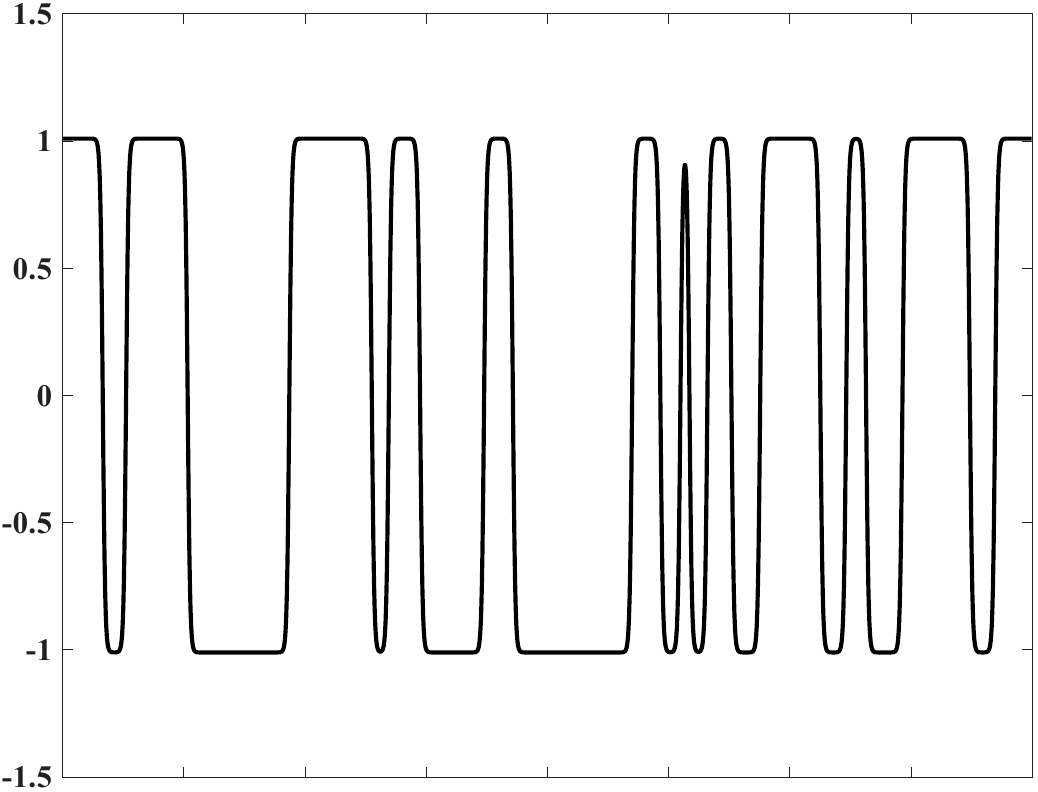}}
    \subfigure[]{\label{domainwall}\includegraphics[width=0.54\columnwidth]{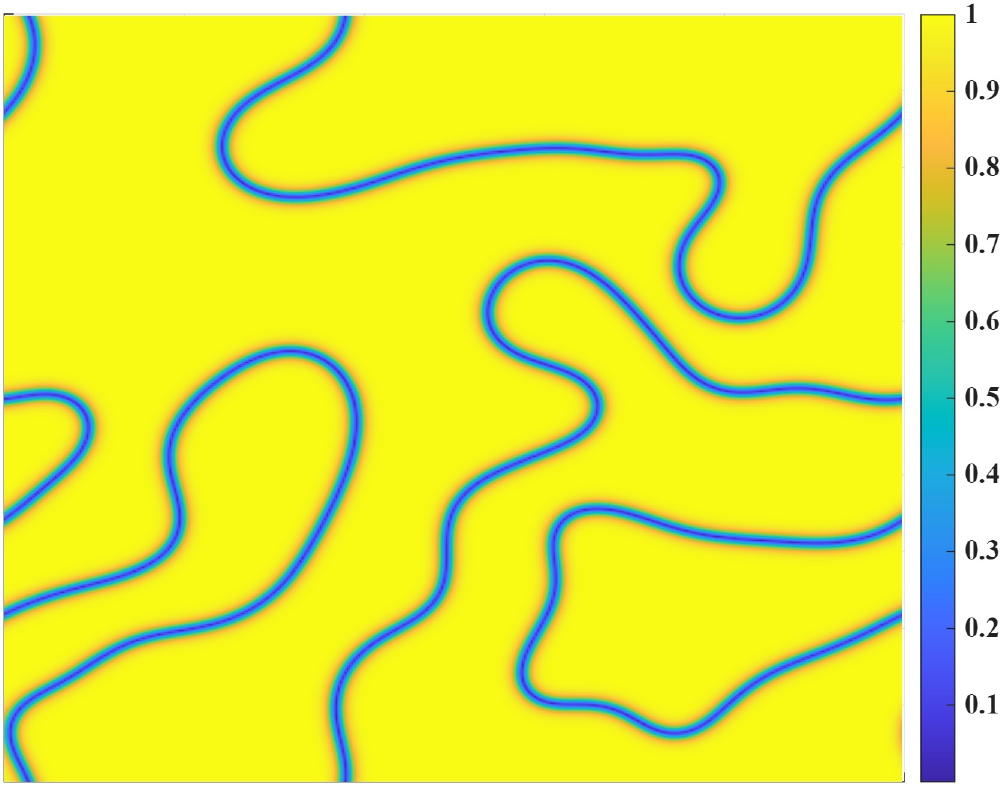}}  
    \subfigure[]{\label{vortex}\includegraphics[width=0.54\columnwidth]{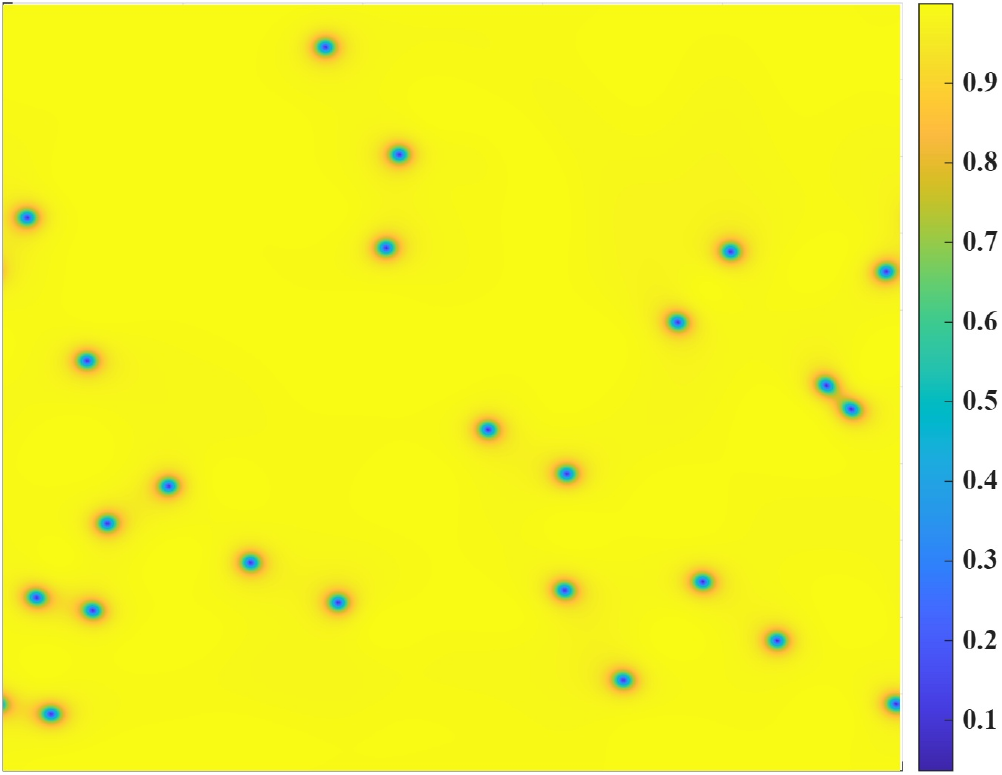}  }
    \subfigure[]{\label{domainwall3D}\includegraphics[width=0.51\columnwidth]{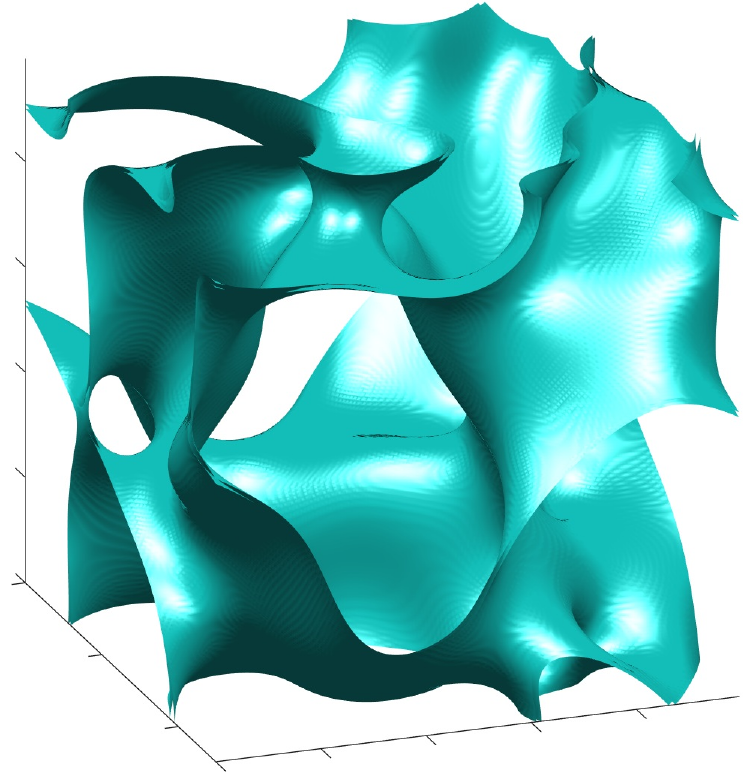}}  
    \subfigure[]{\label{vortexline}\includegraphics[width=0.51\columnwidth]{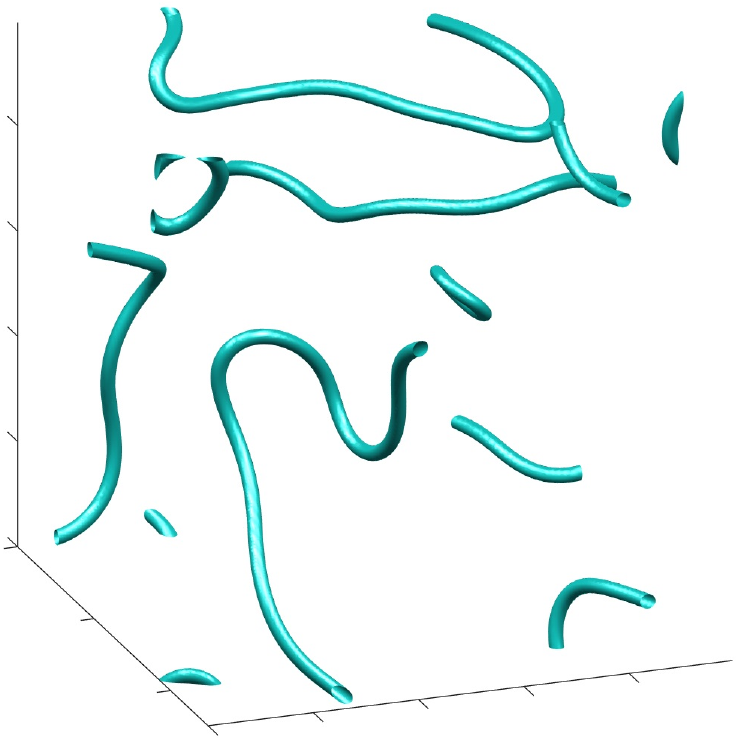} }
    \subfigure[]{\label{monopole}\includegraphics[width=0.56\columnwidth]{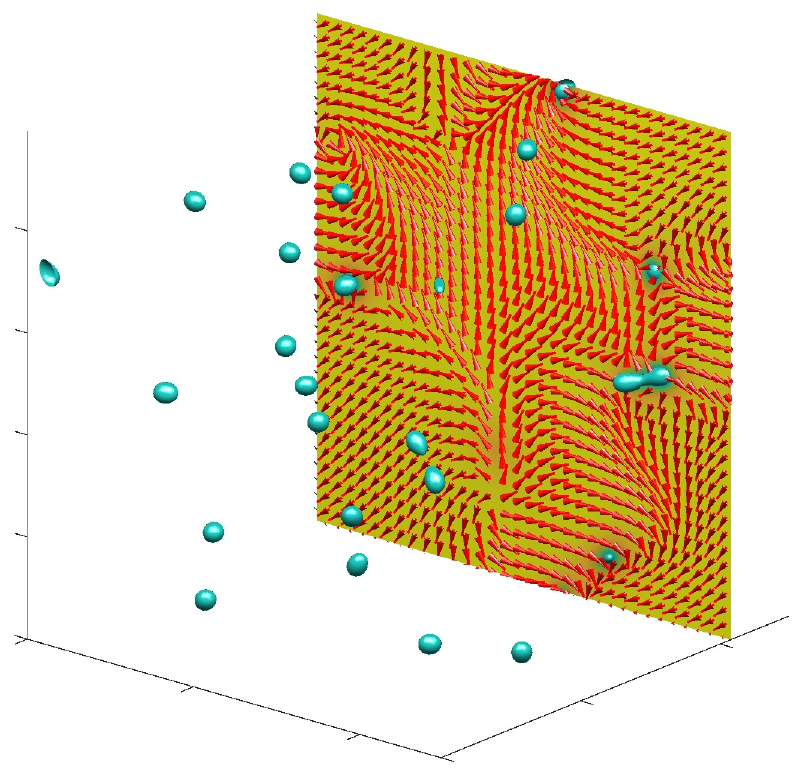} }
    \caption{Spatial configurations of topological defects of different types and dimensions. (a) Kink ($\mathbb{Z}_2$ symmetry). (b) Domain wall ($\mathbb{Z}_2$ symmetry). (c) Vortex ($U(1)$ symmetry). (d) 3D domain wall ($\mathbb{Z}_2$ symmetry). (e) Vortex line ($U(1)$ symmetry). (f) Monopole ($O(3)$ symmetry)
    }\label{topologicaldefectexample}
\end{figure*}

Topological defects commonly arise in nonequilibrium phase transitions involving symmetry breaking. When a system is driven through a critical point at a finite rate, different spatial regions independently choose symmetry-broken states. This picture is captured by the Kibble-Zurek (KZ) mechanism \cite{Kibble:1976sj,Zurek:1985qw,Zurek:1996sj,Digal:1998ak,Dodd:1998aan,Carmi:2000zz,delCampo:2013nla,Sonner:2014tca,Li:2021jqk,Zeng:2022hut,delCampo:2022lqd,Xia:2024wfq,Ma:2024srb,Yang:2025bsw,Wang:2025swz,Xia:2026yrj,Shinn:2025cmh,Zeng:2024rwn,delCampo:2025ocj}, which provides a universal description of defect densities and is one of the most thoroughly studied examples to date. In addition, the formation of topological defects is not limited to thermal quenches; quantum quenches, superfluid instabilities, and cosmological phase transitions can also generate defects. Depending on the symmetry, topological defects can be classified into different types, such as kinks and domain walls for $\mathbb{Z}_2$ symmetry, vortices and vortex lines for $U(1)$ symmetry, or monopoles for $O(3)$ symmetry. Fig.~\ref{topologicaldefectexample} shows the shapes of the order parameters of different types of topological defects. To describe the properties of topological defects, some universal scaling laws have been proposed. For example, the KZ scaling describes the relation between the defect density and the quench rate $\tau_Q$. After the topological defects are formed, the correlations between defect interfaces satisfy Porod scaling \cite{PhysRevB.38.2703,PhysRevLett.67.2670,Bray_1994,PhysRevB.49.14958,PURI1992211,PhysRevLett.99.234505,PhysRevLett.130.128101}. In addition, there are the Kolmogorov scaling \cite{10.1098/rspa.1991.0075,JMaurer1998,10.1063_1.3504375,PhysRevLett.94.065302,Adams:2012pj,Adams:2013vsa,PhysRevX.2.041001,Yang:2024hom,Zeng:2024rwn,delCampo:2025ocj}  for superfluid vortex turbulence, and scaling laws describing subsequent defect evolution and annihilation \cite{PhysRevLett.110.228101,PhysRevLett.134.167101,Zeng:2024rwn}, among others.

These universal scaling laws constitute the foundation of the dynamics of topological defects, and many of them have been experimentally verified. However, the above scaling behaviors largely ignore the contribution of the defect cores. Although topological defects differ in different systems and symmetries, the shapes of their cores can all be fitted by a simple tanh function, which takes the form: \begin{equation}
F(r) = f_0 \tanh\left( \frac{r-r_0}{\sqrt{2}\,\xi} \right),
\label{eqhealingfit}
\end{equation}
in which, $r_0$ denotes the position of the defect core and $\xi$ is the healing length. 
This strongly suggests that the core shapes of the order parameters of these topological defects may leave universal imprints in momentum space. Motivated by this, we define a form factor $S_f$, which takes the form: \begin{align}
S_f(k)=\frac{1}{V\Omega_{d-1}}
\int_{S^{d-1}}
d\Omega_{\widehat{\mathbf{k}}}\,\left|\int d^d r\,
(|\psi(\mathbf{r})|-f_0)
e^{-i\mathbf{k}\cdot\mathbf{r}}\right|^2,\label{eqstructureFactor}
\end{align} 
where $V$ is the volume, and $\psi(\mathbf{r})$ is the spatial distribution of the order parameter. By doing so, we find that the form factor $S_f$ of the order parameter of topological defects exhibits a robust scaling behavior of the form $S_f \propto k^{-(d+p+2)}$, where $d$ is the spatial dimension and $p$ is the defect codimension.

The form factor can serve as a universal fingerprint spectrum, because the scaling behavior of $S_f$ depends only on the spatial dimension and the defect codimension. Moreover, existing experimental measurements can directly or indirectly obtain the spatial distribution of the order parameter of topological defects, such as Josephson scanning tunneling microscopy or other diffraction techniques \cite{PhysRevLett.62.214,PhysRevLett.75.2754,Zhou2017,PhysRevLett.80.3606}. Therefore, $S_f$ is in principle experimentally detectable. Numerically, the order parameter at the defect core can be extracted very easily, making it a powerful tool for studying the properties of topological defects.

\section{Results}
{\bf{\emph{Proof of the scaling behavior of $S_f$}}.} 
In this subsection, we present the mathematical derivation of the universal scaling of the form factor. Briefly, the distribution of topological defects in $d$-dimensional space can be decomposed into two dimensions: the geometric dimension $q$ of the defects and the transverse dimension, i.e., the codimension $p = d - q$.
Since the cores of topological defects can all be fitted by Eq.~(\ref{eqhealingfit}), after taking the absolute value of the order parameter, these defects universally exhibit a V-shaped cusp structure. This is the main reason for the scaling behavior of $S_f$ in the large-$k$ regime.
In a locally flat coordinate patch, we can write $\mathbf{r}=(\mathbf{s},\mathbf{y})\in\mathbb{R}^q\times\mathbb{R}^p$ and define the transverse distance
\begin{equation}
\rho=|\mathbf{y}|.
\label{eq:transverse-distance}
\end{equation}
For $k_\perp\to\infty$ and $\mathbf{k}_\perp\neq0$, we have  (see book \cite{GelfandShilov1964} Chap. II, Sec. 3.3)
\begin{align}
\int_{\mathbb{R}^p}
d^p y\,
|\mathbf{y}|
e^{-i\mathbf{k}_\perp\cdot\mathbf{y}}
&=
-A_p k_\perp^{-(p+1)}
+o\!\left(k_\perp^{-(p+1)}\right),\nonumber\\ A_p
&=
2^p\pi^{(p-1)/2}
\Gamma\!\left(\frac{p+1}{2}\right).
\label{eqlocacusptransform}
\end{align}
Let $\phi(\mathbf{r}) \equiv |\psi(\mathbf{r})| - f_0$ be the form fluctuation of an order parameter relative to its uniform background value $f_0$. We work in a finite periodic observation region or with a smooth observation window $W_V$ that equals unity around every resolved core. The window function $W_V$ is introduced to ensure the convergence of the integral, and it takes the value $W_V=1$ within the finite window. 
Near the defect core, assume the distance expansion
\begin{equation}
\phi(\mathbf{s},\mathbf{y})
=
c_0(\mathbf{s})
+c_1(\mathbf{s})\rho
+\mathcal{O}(\rho^2).\label{eq:core-distance-expansion}
\end{equation}
Since $c_0(\mathbf{s})$ does not depend on $\mathbf{y}$, the Fourier transform along the defect direction contributes a finite constant. Even if there is a slight bending, it does not diverge within a finite window. Therefore, the dominant phase in the Fourier transform is $c_1(\mathbf{s})$. Then, we can write
\begin{equation}
g(\mathbf{s})=c_1(\mathbf{s})W_V(\mathbf{s}),
\qquad
\widehat{g}(\boldsymbol{\kappa})
=
\int d^q s\,
g(\mathbf{s})
e^{- i\boldsymbol{\kappa}\cdot\mathbf{s}},
\label{eq:tangential-window}
\end{equation}
therefore
\begin{align}
&\widetilde{\phi}_V
(\mathbf{k}_\parallel,\mathbf{k}_\perp)\nonumber\\
&=\int d^d r\,
W_V(\mathbf{r})\phi(\mathbf{r})
e^{-i\mathbf{k}\cdot\mathbf{r}}\nonumber\\
&=\left(\int d^q r\,
c_1(\mathbf{s})W_V(\mathbf{s})
e^{-i\mathbf{k_\parallel}\cdot \mathbf{s}}\right)\left( \int
d^p y\,
|\mathbf{y}|
e^{-i\mathbf{k}_\perp\cdot\mathbf{y}}
\right)\nonumber\\
&=
-A_p k_\perp^{-(p+1)}
\widehat{g}(\mathbf{k}_\parallel)
+o\!\left(k_\perp^{-(p+1)}\right).
\label{eq:finite-patch-transform}
\end{align}
For $k>0$, the normalized angle-averaged form factor is
\begin{align}
S_f(k)
&=
\frac{1}{V\Omega_{d-1}}
\int_{S^{d-1}}
d\Omega_{\widehat{\mathbf{k}}}\,
\left|
\widetilde{\phi}_V
(\mathbf{k}_\parallel,\mathbf{k}_\perp)
\right|^2,\nonumber\\
\Omega_m
&=
\frac{2\pi^{(m+1)/2}}
{\Gamma\!\left((m+1)/2\right)}.
\label{eq:normalized-structure-factor}
\end{align}
where
\begin{equation}
    d\Omega_{\hat{\mathbf{k}}} = k^{-q}\left[1 + o(1)\right] d\Omega_{p-1}\, d^q k_\parallel.
\end{equation}
We will present the derivation of the angular volume element in the supplementary materials. Combining the above derivation, we obtain
\begin{align}
S_f(k)
&\propto
\frac{A_p^2\Omega_{p-1}(2\pi)^q}
{\Omega_{d-1}V}
\left[
\int d^q s\,
|c_1(\mathbf{s})W_V(\mathbf{s})|^2
\right]
k^{-(d+p+2)}.
\label{eq:general-shell-tail}
\end{align}
In Eq.~(\ref{eq:general-shell-tail}), since we have imposed constraints on the integral term, it is not divergent. Therefore, the dominant contribution to the scaling behavior is given by $k^{-(d+p+2)}$.

{\bf{\emph{Numerical results}}.}
Next, we perform numerical verification of the theory. We will take the simplest time-dependent Ginzburg-Landau (TDGL) theory as an example. The TDGL equation takes the following form:
\begin{align}
\partial_t\psi=-\alpha(t)\psi-\beta\psi|\psi|^2+\gamma\nabla^2\psi.\label{qeofGl}
\end{align}
In the following calculations, we set $\gamma = 1/2$ and $\beta = 1$. The parameter $\alpha$ is used to control the quench process, with the critical point at $\alpha_c = 0$ and $\alpha < 0$ corresponding to the broken phase. Here $\psi$ denotes the order parameter. As an effective theory, the TDGL theory can be used to describe various types of topological defects \cite{Blundell_1994}. In this manuscript, we mainly consider several common types, namely $\mathbb{Z}_2$ (domain walls), $U(1)$ (vortices), and $O(3)$ (monopoles).
We show the configurations of topological defects of different types and different spatial dimensions for TDGL in Fig.~\ref{topologicaldefectexample}.

The entire nonequilibrium process is as follows. We choose the initial quench point as $\alpha_i = 1$, the final quench point as $\alpha_f = -1$, and the quench time $\tau_Q = 1$, corresponding to an ultrafast quench. After the quench, we let the system evolve freely for a period of time, and the final time is set to $t_f = 50$. Once the topological defects have become sufficiently stable, we measure the healing length $\xi$ of the system. Since the system parameters are chosen identically, the healing length $\xi$ is the same for topological defects of different dimensions. We adopt the simplest method to measure it. For the $\mathbb{Z}_2$ system, we fit the kink profile using Eq.~(\ref{eqhealingfit}); for the $U(1)$ system, we fit the vortex profile using the same formula. For monopoles, the core structure is similar to that of a vortex, so the same fitting method applies.
\begin{figure}
    \centering
    \includegraphics[width=1\columnwidth]{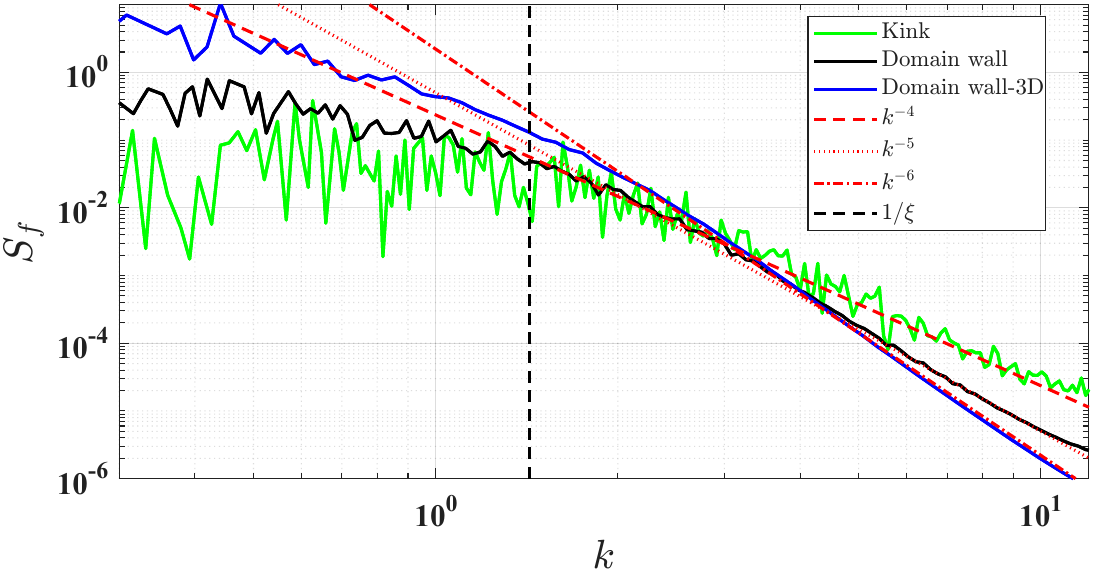}
    \includegraphics[width=1\columnwidth]{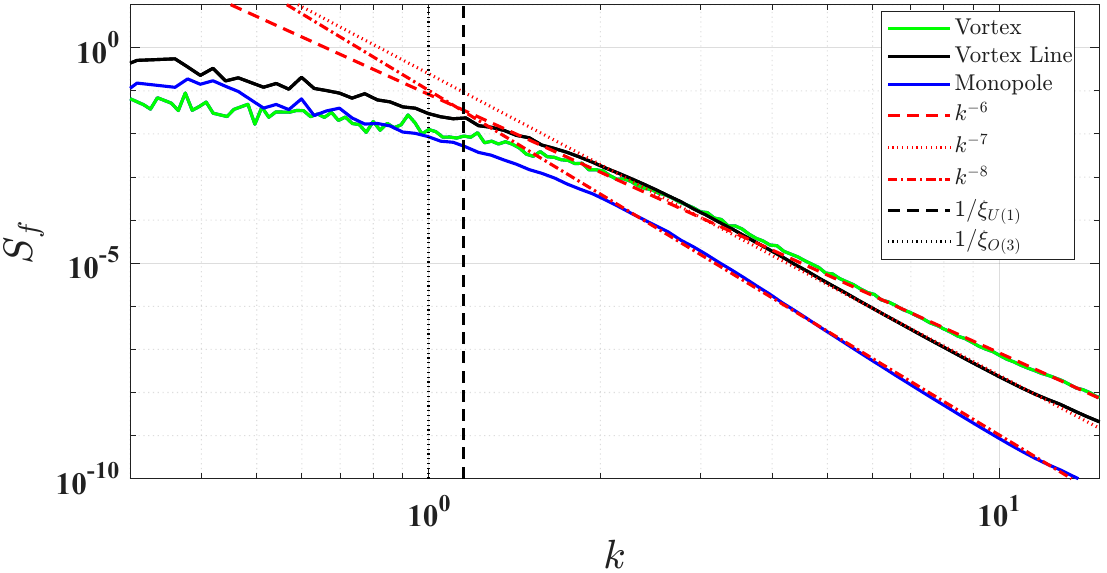}
    \caption{Scaling behavior of topological defects for TDGL model of different types and dimensions. The upper panel corresponds to the system with $\mathbb{Z}_2$ symmetry, and the lower panel corresponds to the system with $U(1)$ and $O(3)$ symmetries.
    }\label{structurefactorexample}
\end{figure}

\begin{figure}
    \centering
    \includegraphics[width=1.15\columnwidth]{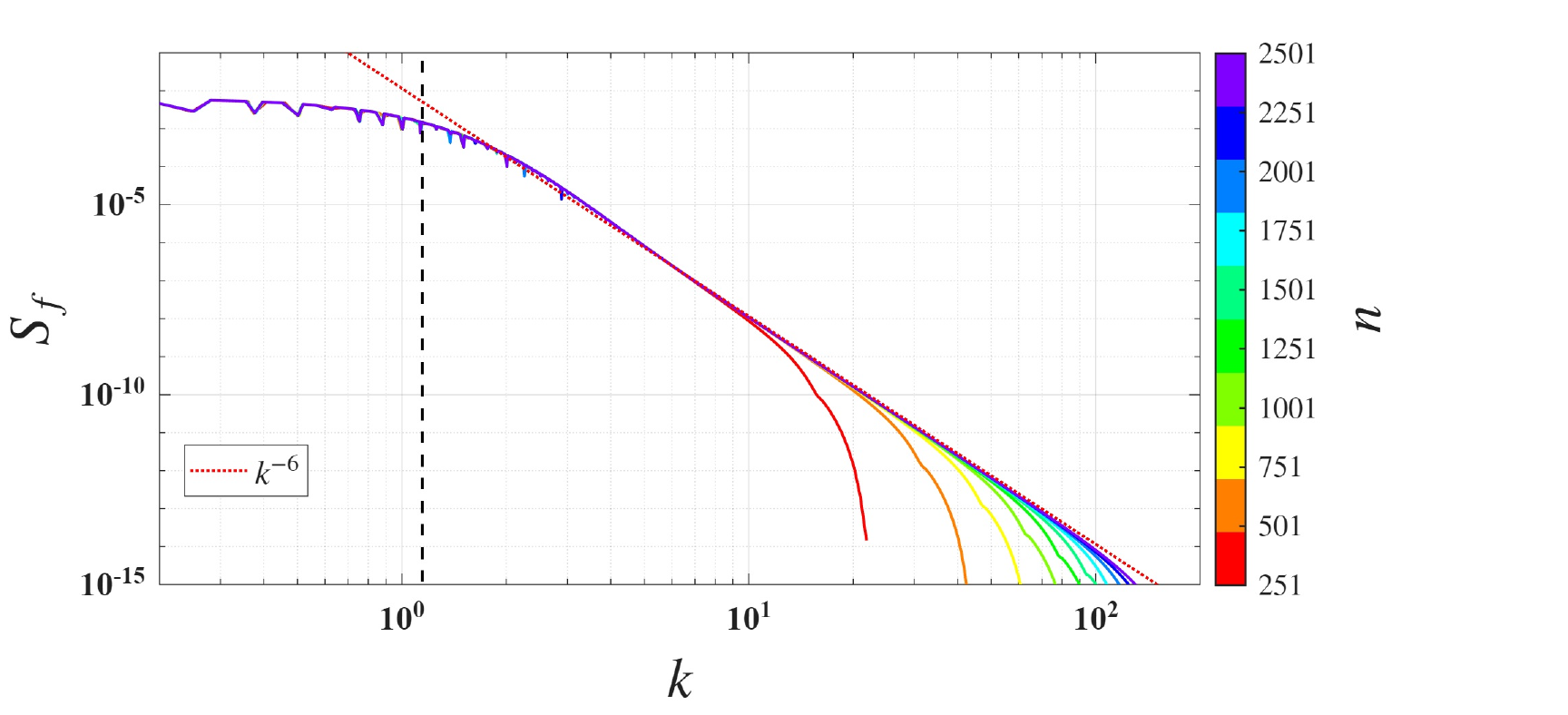}
    \caption{Form factor of single-vortex configurations with different grid numbers, constructed via Eq.~(\ref{eqhealingfit}).
    }\label{OnevortexstructureFactor}
\end{figure}

Next, we compute the form factors of topological defects of different types and dimensions use Eq.~(\ref{eqstructureFactor}), and present all of them in Fig.~\ref{structurefactorexample}. For the form factor of one-dimensional kinks, because the configuration is sufficiently simple, the scaling regime $k>1/\xi$ is well satisfied. However, for higher-dimensional cases, the valid regime shrinks slightly. 
For $k_{\min}$, this may originate from the shape of topological defects and the spatial dimension, which still have some nontrivial contributions in the Fourier integral. The magnitude of $k_{\max}$ mainly depends on the numerical accuracy.
In our calculations, we have uniformly fixed $n = 4L + 1$, which may be quite adequate for one dimension, but may lead to slightly lower precision in the Fourier transform for higher dimensions. As an example, we construct a simple vortex structure using Eq.~(\ref{eqhealingfit}) and display it in Fig.~\ref{OnevortexstructureFactor}, with the same healing length as in Fig.~\ref{structurefactorexample}. We test the configuration with different grid numbers. It can be observed that as the number of grid points increases, the valid regime of the scaling behavior becomes larger. Nevertheless, with the current grid numbers, the form factor still exhibits a well-defined scaling behavior, namely $S_f(k) \sim k^{-(d+p+2)}$.

\begin{table}[h]
\centering
\footnotesize  
\renewcommand{\arraystretch}{0.9} 
\begin{tabular}{|c|c|c|c|c|}
\hline
 & TDGL & GP & Holographic & FRW (PRS) \\
\hline
$\mathbb{Z}_2$-1D & $-4.07\pm0.10$ & -- & $-4.10\pm0.08$ & -- \\
\hline
$\mathbb{Z}_2$-2D & $-5.10\pm0.03$ & -- & $-5.11\pm0.05$ & $-4.99\pm0.04$ \\
\hline
$\mathbb{Z}_2$-3D & $-5.97\pm0.05$ & -- & -- & $-6.02\pm0.04$ \\
\hline
$U(1)$-2D & $-6.01\pm0.04$ & $-6.09\pm0.04$ & $-6.08\pm0.05$ & -- \\
\hline
$U(1)$-3D & $-7.01\pm0.04$ & $-6.99\pm0.04$ & $-7.00\pm0.06$ & -- \\
\hline
$O(3)$-3D & $-8.05\pm0.06$ & -- & -- & -- \\
\hline
\end{tabular}
\caption{Scaling behaviors for different models and dimensions. The empty entries indicate cases that we did not compute. The fitting intervals are presented in the supplementary material.}
\label{tabsum}
\end{table}

To further test the generality of the above conclusions, we extend our calculations to the Gross–Pitaevskii (GP) equation, holographic model, and Klein-Gordon (KG) equations in the Friedmann-Robertson-Walker (FRW) background \cite{Press:1989yh}. The GP model, in addition to dissipation, also possesses wave-like properties, and its relaxation behavior differs significantly from that of the TDGL model. For physics beyond weak coupling, we adopt the holographic model describing strongly coupled superconductors \cite{Hartnoll:2008vx,Sonner:2014tca,Guo:2018mip,Liu_2020,Yang:2023dvk,Lan:2023gyc,Zeng:2024rwn,Xia:2024wfq,Yang:2025bsw,An:2026bcw}. In theory, the physical pictures of strong and weak coupling are drastically different, making this test crucial for establishing the universality of the scaling law. Furthermore, we consider systems with an expanding background, namely the KG equation in the FRW background. Unlike defects in a fixed box in flat spacetime, cosmic expansion leads to a dynamical compression of the comoving healing length $\xi_{\rm com}$ of the defects. To isolate the potential effects of the expanding geometry on the scaling behavior, we consider both the standard FRW evolution and the Press-Ryden-Spergel (PRS) modification \cite{Press:1989yh}, which fixes the comoving thickness. All numerical results are summarized in Table~\ref{tabsum}. The numerical results demonstrate that different systems and different types of topological defects all obey the scaling behavior given by $S_f$. We provide the computational details for the other models in the supplementary material.

{\bf{\emph{Numerical details}}.}
In the spatial directions, we employ Fourier spectral methods. For the holographic model, we adopt the Chebyshev spectral method in the holographic direction. For the time evolution, we use the fourth-order Runge-Kutta method.
For the 1D TDGL model, the spatial scale is $L=400$, and for the 2D case, $L=100$. Since 3D calculations require excessive computational resources, we take $L=50$. The chosen time for the order parameter is $t_f=50$. For the GP model, the values of $L$ in 2D and 3D are the same as in the TDGL model, with $t_f=50$. For the holographic model, we take $L=200$ in 1D, $L=100$ in 2D, and $L=40$ in 3D. The times chosen for different types of topological defects are $t_f=150$ for kinks, $t_f=150$ for domain walls, and $t_f=275$ for vortices. In principle, the FRW evolution requires a considerably larger scale; however, since our aim is only to generate topological defects, we choose $L=50$.

\section{Conclusion}
In this paper, we have defined a form factor $S_f$ and investigated its universal scaling behavior for topological defects in different systems and with different symmetries. In the regime $k > 1/\xi$, a well-defined scaling behavior is found, taking the form $S_f \propto k^{-(d+p+2)}$, where $d$ is the spatial dimension and $p$ is the defect codimension.

This scaling behavior is fundamentally different from conventional scaling laws. Previously, attention was paid to correlations between defects, whereas we are more concerned with the imprint of defect cores in momentum space. Our work is more like studying the fingerprints of topological defects, meaning that this work can be regarded as a supplement to many scaling behaviors in the large-$k$ regime, such as Porod scaling, whose valid regime is $1/L(t) < k < 1/\xi$.

There remain several open questions. In this paper, we have only calculated several typical types of topological defects. Whether other symmetries also obey this scaling behavior remains to be examined. A more comprehensive verification can be carried out in future work. For systems with coexisting multiple order parameters that breaks different symmetry, $S_f$ could in principle provide a quantitative indicator for determining which order parameter dominates the evolution, but this has not been investigated in the present work. We plan to address these issues in future studies.

{\emph{Acknowledgement}}.---ZQZ thanks Ji-Yu Song for useful discussions. This work is supported by the National Natural Science Foundation of China (grant nos. 12575054, 12533001, 12575049, 12473001, 12205039, 12305058, and 11965013). ZYN is partially supported by Yunnan High-level Talent Training Support Plan Young $\&$ Elite Talents Project (grant no. YNWR-QNBJ-2018-181). This work is also supported by the National SKA Program of China (grant nos. 2022SKA0110200 and 2022SKA0110203) and the 111 Project (grant no. B16009).

\bibliographystyle{apsrev4-1}
\bibliography{reference}

@article{Kibble:1976sj,
    author = "Kibble, T. W. B.",
    title = "{Topology of Cosmic Domains and Strings}",
    reportNumber = "ICTP/75/5",
    doi = "10.1088/0305-4470/9/8/029",
    journal = "J. Phys. A",
    volume = "9",
    pages = "1387--1398",
    year = "1976"
}

@article{Zurek:1985qw,
    author = "Zurek, W. H.",
    title = "{Cosmological Experiments in Superfluid Helium?}",
    doi = "10.1038/317505a0",
    journal = "Nature",
    volume = "317",
    pages = "505--508",
    year = "1985"
}

@article{Zurek:1996sj,
    author = "Zurek, W. H.",
    title = "{Cosmological experiments in condensed matter systems}",
    eprint = "cond-mat/9607135",
    archivePrefix = "arXiv",
    reportNumber = "LA-UR-95-2269, LAUR-95-2269-(LOS-ALAMOS)",
    doi = "10.1016/S0370-1573(96)00009-9",
    journal = "Phys. Rept.",
    volume = "276",
    pages = "177--221",
    year = "1996"
}

@article{delCampo:2013nla,
    author = "del Campo, Adolfo and Zurek, Wojciech H.",
    editor = "Gauntlett, Jerome",
    title = "{Universality of phase transition dynamics: Topological Defects from Symmetry Breaking}",
    eprint = "1310.1600",
    archivePrefix = "arXiv",
    primaryClass = "cond-mat.stat-mech",
    reportNumber = "LA-UR-13-27755, LA-UR-13-27755",
    doi = "10.1142/S0217751X1430018X",
    journal = "Int. J. Mod. Phys. A",
    volume = "29",
    number = "8",
    pages = "1430018",
    year = "2014"
}

@article{Sonner:2014tca,
    author = "Sonner, Julian and del Campo, Adolfo and Zurek, Wojciech H.",
    title = "{Universal far-from-equilibrium Dynamics of a Holographic Superconductor}",
    eprint = "1406.2329",
    archivePrefix = "arXiv",
    primaryClass = "hep-th",
    reportNumber = "MIT-CTP-4553, LA-UR-14-24054",
    doi = "10.1038/ncomms8406",
    journal = "Nature Commun.",
    volume = "6",
    pages = "7406",
    year = "2015"
}

@article{Zeng:2022hut,
    author = "Zeng, Hua-Bi and Xia, Chuan-Yin and del Campo, Adolfo",
    title = "{Universal Breakdown of Kibble-Zurek Scaling in Fast Quenches across a Phase Transition}",
    eprint = "2204.13529",
    archivePrefix = "arXiv",
    primaryClass = "cond-mat.stat-mech",
    doi = "10.1103/PhysRevLett.130.060402",
    journal = "Phys. Rev. Lett.",
    volume = "130",
    number = "6",
    pages = "060402",
    year = "2023"
}

@article{Xia:2024wfq,
    author = "Xia, Chuan-Yin and Zeng, Hua-Bi and Grabarits, Andr{\'a}s and del Campo, Adolfo",
    title = "{Kibble-Zurek mechanism and beyond in a holographic superfluid disk}",
    eprint = "2406.09433",
    archivePrefix = "arXiv",
    primaryClass = "cond-mat.stat-mech",
    doi = "10.1038/s41467-026-69940-w",
    journal = "Nature Commun.",
    volume = "17",
    number = "1",
    pages = "3668",
    year = "2026"
}

@article{Yang:2025bsw,
    author = "Yang, Peng and Xia, Chuan-Yin and Grieninger, Sebastian and Zeng, Hua-Bi and Baggioli, Matteo",
    title = "{Topological Defect Formation beyond the Kibble-Zurek Mechanism in Crossover Transitions with Approximate Symmetries}",
    eprint = "2508.05964",
    archivePrefix = "arXiv",
    primaryClass = "cond-mat.stat-mech",
    doi = "10.1103/clvs-yk7v",
    journal = "Phys. Rev. Lett.",
    volume = "136",
    number = "5",
    pages = "051602",
    year = "2026"
}

@article{Hartnoll:2008vx,
    author = "Hartnoll, Sean A. and Herzog, Christopher P. and Horowitz, Gary T.",
    title = "{Building a Holographic Superconductor}",
    eprint = "0803.3295",
    archivePrefix = "arXiv",
    primaryClass = "hep-th",
    reportNumber = "NSF-KITP-08-38, PUPT-2261",
    doi = "10.1103/PhysRevLett.101.031601",
    journal = "Phys. Rev. Lett.",
    volume = "101",
    pages = "031601",
    year = "2008"
}

@article{Ma:2024srb,
    author = "Ma, Tian-Chi and Shi, Han-Qing and Zhang, Hai-Qing and del Campo, Adolfo",
    title = "{Universal critical holography and domain wall formation}",
    eprint = "2406.05167",
    archivePrefix = "arXiv",
    primaryClass = "cond-mat.stat-mech",
    doi = "10.1103/PhysRevResearch.7.013096",
    journal = "Phys. Rev. Res.",
    volume = "7",
    number = "1",
    pages = "013096",
    year = "2025"
}

@article{delCampo:2022lqd,
    author = "del Campo, Adolfo and G{\'o}mez-Ruiz, Fernando Javier and Zhang, Hai-Qing",
    title = "{Locality of spontaneous symmetry breaking and universal spacing distribution of topological defects formed across a phase transition}",
    eprint = "2202.11731",
    archivePrefix = "arXiv",
    primaryClass = "cond-mat.stat-mech",
    doi = "10.1103/PhysRevB.106.L140101",
    journal = "Phys. Rev. B",
    volume = "106",
    number = "14",
    pages = "L140101",
    year = "2022"
}

@article{Li:2021jqk,
    author = "Li, Zhi-Hong and Shi, Han-Qing and Zhang, Hai-Qing",
    title = "{Holographic topological defects in a ring: role of diverse boundary conditions}",
    eprint = "2111.15230",
    archivePrefix = "arXiv",
    primaryClass = "hep-th",
    doi = "10.1007/JHEP05(2022)056",
    journal = "JHEP",
    volume = "05",
    pages = "056",
    year = "2022"
}

@article{Wang:2025swz,
    author = "Wang, Chen-Xu and Grabarits, Andr{\'a}s and Cui, Jin-Ming and Zeng, Hua-Bi and Huang, Yun-Feng and Li, Chuan-Feng and del Campo, Adolfo",
    title = "{Quantum Quenches from the Critical Point: Theory and Experimental Validation in a Trapped-Ion Quantum Simulator}",
    eprint = "2507.01087",
    archivePrefix = "arXiv",
    primaryClass = "quant-ph",
    month = "7",
    year = "2025"
}

@article{Digal:1998ak,
    author = "Digal, Sanatan and Ray, Rajarshi and Srivastava, Ajit M.",
    title = "{Observing correlated production of defect - anti-defects in liquid crystals}",
    eprint = "hep-ph/9805502",
    archivePrefix = "arXiv",
    reportNumber = "IP-BBSR-98-9",
    doi = "10.1103/PhysRevLett.83.5030",
    journal = "Phys. Rev. Lett.",
    volume = "83",
    pages = "5030",
    year = "1999"
}

@article{Dodd:1998aan,
    author = "Dodd, M. E. and Hendry, P. C. and Lawson, N. S. and McClintock, P. V. E. and Williams, C. D. H.",
    title = "{Nonappearance of Vortices in Fast Mechanical Expansions of Liquid 4He through the Lambda Transition}",
    eprint = "cond-mat/9808117",
    archivePrefix = "arXiv",
    reportNumber = "CW980810-2",
    doi = "10.1103/PhysRevLett.81.3703",
    journal = "Phys. Rev. Lett.",
    volume = "81",
    number = "17",
    pages = "3703--3706",
    year = "1998"
}

@article{Carmi:2000zz,
    author = "Carmi, Raz and Polturak, Emil and Koren, Gad",
    title = "{Observation of Spontaneous Flux Generation in a Multi-Josephson-Junction Loop}",
    doi = "10.1103/PhysRevLett.84.4966",
    journal = "Phys. Rev. Lett.",
    volume = "84",
    pages = "4966--4969",
    year = "2000"
}

@article{Xia:2026yrj,
    author = "Xia, Chuan-Yin and Grabarits, Andr{\'a}s and Zeng, Hua-Bi and del Campo, Adolfo",
    title = "{Evolution of Vortex Strings after a Thermal Quench in a Holographic Superfluid}",
    eprint = "2601.14328",
    archivePrefix = "arXiv",
    primaryClass = "hep-th",
    month = "1",
    year = "2026"
}

@article{PhysRevB.38.2703,
  title = {Dynamical scaling, domain-growth kinetics, and domain-wall shapes of quenched two-dimensional anisotropic XY models},
  author = {Mouritsen, Ole G. and Praestgaard, Eigil},
  journal = {Phys. Rev. B},
  volume = {38},
  issue = {4},
  pages = {2703--2714},
  numpages = {0},
  year = {1988},
  month = {Aug},
  publisher = {American Physical Society},
  doi = {10.1103/PhysRevB.38.2703},
  url = {https://link.aps.org/doi/10.1103/PhysRevB.38.2703}
}

@article{Bray_1994,
   title={Theory of phase-ordering kinetics},
   volume={43},
   ISSN={1460-6976},
   url={http://dx.doi.org/10.1080/00018739400101505},
   DOI={10.1080/00018739400101505},
   number={3},
   journal={Advances in Physics},
   publisher={Informa UK Limited},
   author={Bray, A.J.},
   year={1994},
   month=June, pages={357–459} }

@article{PhysRevB.49.14958,
  title = {Computer simulation of vapor-liquid phase separation in two- and three-dimensional fluids: Growth law of domain size},
  author = {Yamamoto, R. and Nakanishi, K.},
  journal = {Phys. Rev. B},
  volume = {49},
  issue = {21},
  pages = {14958--14966},
  numpages = {0},
  year = {1994},
  month = {Jun},
  publisher = {American Physical Society},
  doi = {10.1103/PhysRevB.49.14958},
  url = {https://link.aps.org/doi/10.1103/PhysRevB.49.14958}
}

@article{PURI1992211,
title = {Asymptotic structure factor for the two-component Ginzburg-Landau equation},
journal = {Physics Letters A},
volume = {164},
number = {2},
pages = {211-217},
year = {1992},
issn = {0375-9601},
doi = {https://doi.org/10.1016/0375-9601(92)90705-Q},
url = {https://www.sciencedirect.com/science/article/pii/037596019290705Q},
author = {Sanjay Puri}
}

@article{Yang:2024hom,
    author = "Yang, Wei-Can and Xia, Chuan-Yin and Tian, Yu and Tsubota, Makoto and Zeng, Hua-Bi",
    title = "{Emergence of Large-Scale Structures in Holographic Superfluid Turbulence}",
    eprint = "2402.17980",
    archivePrefix = "arXiv",
    primaryClass = "hep-th",
    month = "2",
    year = "2024"
}

@article{Zeng:2024rwn,
    author = "Zeng, Hua-Bi and Xia, Chuan-Yin and Yang, Wei-Can and Tian, Yu and Tsubota, Makoto",
    title = "{Dissipation and Decay of Three-Dimensional Holographic Quantum Turbulence}",
    eprint = "2408.13620",
    archivePrefix = "arXiv",
    primaryClass = "hep-th",
    doi = "10.1103/PhysRevLett.134.091603",
    journal = "Phys. Rev. Lett.",
    volume = "134",
    number = "9",
    pages = "091603",
    year = "2025"
}

@article{10.1098/rspa.1991.0075,
    author = {Kolmogorov, Andrei Nikolaevich and Levin, V. and Hunt, Julian Charles Roland and Phillips, Owen Martin and Williams, David},
    title = {The local structure of turbulence in incompressible viscous fluid for very large Reynolds numbers},
    journal = {Proceedings of the Royal Society of London. Series A: Mathematical and Physical Sciences},
    volume = {434},
    number = {1890},
    pages = {9-13},
    year = {1991},
    month = {07},
    issn = {0962-8444},
    doi = {10.1098/rspa.1991.0075},
    url = {https://doi.org/10.1098/rspa.1991.0075},
}

@article{JMaurer1998,
    year = {1998},
    month = {jul},
    publisher = {},
    volume = {43},
    number = {1},
    pages = {29},
    author = {J. Maurer and P. Tabeling},
    title = {Local investigation of superfluid turbulence},
    journal = {Europhysics Letters}
    }

@article{10.1063_1.3504375,
    author = {Salort, J. and Baudet, C. and Castaing, B. and Chabaud, B. and Daviaud, F. and Didelot, T. and Diribarne, P. and Dubrulle, B. and Gagne, Y. and Gauthier, F. and Girard, A. and Hébral, B. and Rousset, B. and Thibault, P. and Roche, P.-E.},
    title = {Turbulent velocity spectra in superfluid flows},
    journal = {Physics of Fluids},
    volume = {22},
    number = {12},
    pages = {125102},
    year = {2010},
    month = {12},
    issn = {1070-6631},
    doi = {10.1063/1.3504375},
    url = {https://doi.org/10.1063/1.3504375}
}

@article{PhysRevLett.94.065302,
  title = {Kolmogorov Spectrum of Superfluid Turbulence: Numerical Analysis of the Gross-Pitaevskii Equation with a Small-Scale Dissipation},
  author = {Kobayashi, Michikazu and Tsubota, Makoto},
  journal = {Phys. Rev. Lett.},
  volume = {94},
  issue = {6},
  pages = {065302},
  numpages = {4},
  year = {2005},
  month = {Feb},
  publisher = {American Physical Society},
  doi = {10.1103/PhysRevLett.94.065302},
  url = {https://link.aps.org/doi/10.1103/PhysRevLett.94.065302}
}

@article{PhysRevX.2.041001,
  title = {Energy Spectra of Vortex Distributions in Two-Dimensional Quantum Turbulence},
  author = {Bradley, Ashton S. and Anderson, Brian P.},
  journal = {Phys. Rev. X},
  volume = {2},
  issue = {4},
  pages = {041001},
  numpages = {20},
  year = {2012},
  month = {Oct},
  publisher = {American Physical Society},
  doi = {10.1103/PhysRevX.2.041001},
  url = {https://link.aps.org/doi/10.1103/PhysRevX.2.041001}
}

@book{GelfandShilov1964,
  author    = {I. M. Gel'fand and G. E. Shilov},
  title     = {Generalized Functions},
  volume    = {1},
  subtitle  = {Properties and Operations},
  publisher = {Academic Press},
  address   = {New York},
  year      = {1964}
}

@article{Shinn:2025cmh,
    author = "Shinn, Seong-Ho and Massaro, Matteo and Thudiyangal, Mithun and del Campo, Adolfo",
    title = "{Spontaneous Quantum Turbulence in a Newborn Bose-Einstein Condensate via the Kibble-Zurek Mechanism}",
    eprint = "2506.21670",
    archivePrefix = "arXiv",
    primaryClass = "cond-mat.quant-gas",
    doi = "10.1103/b16v-hwp2",
    journal = "Phys. Rev. Lett.",
    volume = "137",
    number = "2",
    pages = "020402",
    year = "2026"
}

@article{PhysRevLett.67.2670,
  title = {Asymptotic structure factor and power-law tails for phase ordering in systems with continuous symmetry},
  author = {Bray, A. J. and Puri, Sanjay},
  journal = {Phys. Rev. Lett.},
  volume = {67},
  issue = {19},
  pages = {2670--2673},
  numpages = {0},
  year = {1991},
  month = {Nov},
  publisher = {American Physical Society},
  doi = {10.1103/PhysRevLett.67.2670},
  url = {https://link.aps.org/doi/10.1103/PhysRevLett.67.2670}
}

@article{PhysRevLett.130.128101,
  title = {Defect Line Coarsening and Refinement in Active Nematics},
  author = {Kralj, Nika and Ravnik, Miha and Kos, \ifmmode \check{Z}\else \v{Z}\fi{}iga},
  journal = {Phys. Rev. Lett.},
  volume = {130},
  issue = {12},
  pages = {128101},
  numpages = {6},
  year = {2023},
  month = {Mar},
  publisher = {American Physical Society},
  doi = {10.1103/PhysRevLett.130.128101},
  url = {https://link.aps.org/doi/10.1103/PhysRevLett.130.128101}
}

@article{PhysRevLett.99.234505,
  title = {Violation of the Porod Law in a Freely Cooling Granular Gas in One Dimension},
  author = {Shinde, Mahendra and Das, Dibyendu and Rajesh, R.},
  journal = {Phys. Rev. Lett.},
  volume = {99},
  issue = {23},
  pages = {234505},
  numpages = {4},
  year = {2007},
  month = {Dec},
  publisher = {American Physical Society},
  doi = {10.1103/PhysRevLett.99.234505},
  url = {https://link.aps.org/doi/10.1103/PhysRevLett.99.234505}
}

@article{PhysRevLett.110.228101,
  title = {Defect Annihilation and Proliferation in Active Nematics},
  author = {Giomi, Luca and Bowick, Mark J. and Ma, Xu and Marchetti, M. Cristina},
  journal = {Phys. Rev. Lett.},
  volume = {110},
  issue = {22},
  pages = {228101},
  numpages = {5},
  year = {2013},
  month = {May},
  publisher = {American Physical Society},
  doi = {10.1103/PhysRevLett.110.228101},
  url = {https://link.aps.org/doi/10.1103/PhysRevLett.110.228101}
}

@article{PhysRevLett.134.167101,
  title = {Nonreciprocal Interactions Reshape Topological Defect Annihilation},
  author = {Rouzaire, Ylann and Pearce, Daniel J. G. and Pagonabarraga, Ignacio and Levis, Demian},
  journal = {Phys. Rev. Lett.},
  volume = {134},
  issue = {16},
  pages = {167101},
  numpages = {6},
  year = {2025},
  month = {Apr},
  publisher = {American Physical Society},
  doi = {10.1103/PhysRevLett.134.167101},
  url = {https://link.aps.org/doi/10.1103/PhysRevLett.134.167101}
}

@article{PhysRevLett.80.3606,
  title = {Observation of the Low Temperature Pseudogap in the Vortex Cores of ${\mathrm{Bi}}_{2}{\mathrm{Sr}}_{2}{\mathrm{CaCu}}_{2}{O}_{8+\ensuremath{\delta}}$},
  author = {Renner, Ch. and Revaz, B. and Kadowaki, K. and Maggio-Aprile, I. and Fischer, \O{}.},
  journal = {Phys. Rev. Lett.},
  volume = {80},
  issue = {16},
  pages = {3606--3609},
  numpages = {0},
  year = {1998},
  month = {Apr},
  publisher = {American Physical Society},
  doi = {10.1103/PhysRevLett.80.3606},
  url = {https://link.aps.org/doi/10.1103/PhysRevLett.80.3606}
}

@article{Liu_2020,
doi = {10.1088/1361-6633/ab4f91},
url = {https://doi.org/10.1088/1361-6633/ab4f91},
year = {2019},
month = {dec},
publisher = {IOP Publishing},
volume = {83},
number = {1},
pages = {016001},
author = {Liu, Hong and Sonner, Julian},
title = {Holographic systems far from equilibrium: a review},
journal = {Reports on Progress in Physics}
}

@article{Lan:2023gyc,
    author = "Lan, Shanquan and Li, Xin and Tian, Yu and Yang, Peng and Zhang, Hongbao",
    title = "{Heating Up Quadruply Quantized Vortices: Splitting Patterns and Dynamical Transitions}",
    eprint = "2311.01316",
    archivePrefix = "arXiv",
    primaryClass = "cond-mat.quant-gas",
    doi = "10.1103/PhysRevLett.131.221602",
    journal = "Phys. Rev. Lett.",
    volume = "131",
    number = "22",
    pages = "221602",
    year = "2023"
}

@article{Yang:2023dvk,
    author = "Yang, Peng and Baggioli, Matteo and Cai, Zi and Tian, Yu and Zhang, Hongbao",
    title = "{Holographic Dissipative Spacetime Supersolids}",
    eprint = "2304.02534",
    archivePrefix = "arXiv",
    primaryClass = "hep-th",
    doi = "10.1103/PhysRevLett.131.221601",
    journal = "Phys. Rev. Lett.",
    volume = "131",
    number = "22",
    pages = "221601",
    year = "2023"
}

@article{Guo:2018mip,
    author = "Guo, Minyong and Keski-Vakkuri, Esko and Liu, Hong and Tian, Yu and Zhang, Hongbao",
    title = "{Dynamical Phase Transition from Nonequilibrium Dynamics of Dark Solitons}",
    eprint = "1810.11424",
    archivePrefix = "arXiv",
    primaryClass = "hep-th",
    reportNumber = "MIT-CTP/5077",
    doi = "10.1103/PhysRevLett.124.031601",
    journal = "Phys. Rev. Lett.",
    volume = "124",
    number = "3",
    pages = "031601",
    year = "2020"
}

@article{Press:1989yh,
    author = "Press, William H. and Ryden, Barbara S. and Spergel, David N.",
    title = "{Dynamical Evolution of Domain Walls in an Expanding Universe}",
    reportNumber = "NSF-ITP-89-51, CFA-1870",
    doi = "10.1086/168151",
    journal = "Astrophys. J.",
    volume = "347",
    pages = "590--604",
    year = "1989"
}

@article{delCampo:2025ocj,
    author = "del Campo, Adolfo and Grabarits, Andr{\'a}s and Makarov, Dmitrii E. and Shinn, Seong-Ho",
    title = "{Quantum Transition Rates in Arbitrary Physical Processes}",
    eprint = "2506.21672",
    archivePrefix = "arXiv",
    primaryClass = "quant-ph",
    doi = "10.1103/l7dq-n61h",
    journal = "Phys. Rev. Lett.",
    volume = "136",
    number = "21",
    pages = "210202",
    year = "2026"
}

@article{Adams:2013vsa,
    author = "Adams, Allan and Chesler, Paul M. and Liu, Hong",
    title = "{Holographic turbulence}",
    eprint = "1307.7267",
    archivePrefix = "arXiv",
    primaryClass = "hep-th",
    reportNumber = "MIT-CTP-4460",
    doi = "10.1103/PhysRevLett.112.151602",
    journal = "Phys. Rev. Lett.",
    volume = "112",
    number = "15",
    pages = "151602",
    year = "2014"
}

@article{Adams:2012pj,
    author = "Adams, Allan and Chesler, Paul M. and Liu, Hong",
    title = "{Holographic Vortex Liquids and Superfluid Turbulence}",
    eprint = "1212.0281",
    archivePrefix = "arXiv",
    primaryClass = "hep-th",
    doi = "10.1126/science.1233529",
    journal = "Science",
    volume = "341",
    pages = "368--372",
    year = "2013"
}

@article{An:2026bcw,
    author = "An, Yu-Ping and Ding, Peng-Bo and Jin, Zhen-Han and Li, Li",
    title = "{Statistically Steady Holographic Quantum Turbulence: Hyperuniform Vortex Matter and Crossover}",
    eprint = "2608.17012",
    archivePrefix = "arXiv",
    primaryClass = "hep-th",
    month = "8",
    year = "2026"
}

@article{Zhou2017,
  author    = {Zhou, Shuang and Shiyanovskii, Sergij V. and Park, Heung-Shik and Lavrentovich, Oleg D.},
  title     = {Fine structure of the topological defect cores studied for disclinations in lyotropic chromonic liquid crystals},
  journal   = {Nature Communications},
  year      = {2017},
  volume    = {8},
  number    = {1},
  pages     = {14974},
  month     = {apr},
  doi       = {10.1038/ncomms14974},
  issn      = {2041-1723},
  publisher = {Nature Publishing Group}
}

@article{Blundell_1994,
   title={Phase-ordering dynamics of the O(<i>n</i>) model: Exact predictions and numerical results},
   volume={49},
   ISSN={1095-3787},
   url={http://dx.doi.org/10.1103/PhysRevE.49.4925},
   DOI={10.1103/physreve.49.4925},
   number={6},
   journal={Physical Review E},
   publisher={American Physical Society (APS)},
   author={Blundell, R. E. and Bray, A. J.},
   year={1994},
   month=June, pages={4925–4937} }

@article{PhysRevLett.75.2754,
  title = {Direct Vortex Lattice Imaging and Tunneling Spectroscopy of Flux Lines on ${\mathrm{YBa}}_{2}{\mathrm{Cu}}_{3}{O}_{7\ensuremath{-}\mathit{\ensuremath{\delta}}}$},
  author = {Maggio-Aprile, I. and Renner, Ch. and Erb, A. and Walker, E. and Fischer, \O{}.},
  journal = {Phys. Rev. Lett.},
  volume = {75},
  issue = {14},
  pages = {2754--2757},
  numpages = {0},
  year = {1995},
  month = {Oct},
  publisher = {American Physical Society},
  doi = {10.1103/PhysRevLett.75.2754},
  url = {https://link.aps.org/doi/10.1103/PhysRevLett.75.2754}
}

@article{PhysRevLett.62.214,
  title = {Scanning-Tunneling-Microscope Observation of the Abrikosov Flux Lattice and the Density of States near and inside a Fluxoid},
  author = {Hess, H. F. and Robinson, R. B. and Dynes, R. C. and Valles, J. M. and Waszczak, J. V.},
  journal = {Phys. Rev. Lett.},
  volume = {62},
  issue = {2},
  pages = {214--216},
  numpages = {0},
  year = {1989},
  month = {Jan},
  publisher = {American Physical Society},
  doi = {10.1103/PhysRevLett.62.214},
  url = {https://link.aps.org/doi/10.1103/PhysRevLett.62.214}
}

\clearpage
\onecolumngrid
\begin{appendix}

\begin{center}

\renewcommand\thefigure{S\arabic{figure}}    
\setcounter{figure}{0} 
\renewcommand{\theequation}{S\arabic{equation}}
\setcounter{equation}{0}
\renewcommand{\thesubsection}{SI\arabic{subsection}}

\end{center}
\setcounter{equation}{0}
\setcounter{table}{0}

\begin{center}
    \textbf{Supplementary Materials}
\end{center}

\subsection{Angular measure}
In this section, we derive the angular measure used in the main text. In d-dimensional spherical coordinates, the volume element is
\begin{equation}
    d^d k = k^{d-1}\, dk \, d\Omega_{\hat{\mathbf{k}}},\label{AM1}
\end{equation}
which can be decomposed into
\begin{equation}
    d^d k = d^q k_\parallel \, d^p k_\perp
    = d^q k_\parallel \cdot k_\perp^{p-1}\, dk_\perp \, d\Omega_{p-1}.\label{AM2}
\end{equation}
For $k^2 = k_\parallel^2 + k_\perp^2$, with $\mathbf{k}_\parallel$ fixed, we have
\begin{equation}
    k\, dk = k_\perp\, dk_\perp \quad \Rightarrow \quad dk_\perp = \frac{k}{k_\perp}\, dk.\label{AM3}
\end{equation}
After substituting Eq.~(\ref{AM1}) and Eq.~(\ref{AM3}) into Eq.~(\ref{AM2}), we obtain
\begin{equation}
    k^{d-1}\, dk \, d\Omega_{\hat{\mathbf{k}}}
    = k \cdot k_\perp^{p-2} \, dk \, d\Omega_{p-1} \, d^q k_\parallel.
\end{equation}
Canceling $dk$ and using $d = p + q$, we have
\begin{align}
    d\Omega_{\hat{\mathbf{k}}}
    &= k^{-(p+q-2)} \cdot k_\perp^{p-2} \, d\Omega_{p-1} \, d^q k_\parallel= \left(\frac{k_\perp}{k}\right)^{p-2} k^{-q} \, d\Omega_{p-1} \, d^q k_\parallel.
\end{align}
Now, in the asymptotic regime $k \to \infty$ with $\mathbf{k}_\parallel$ fixed (or $k_\parallel \ll k$):
\begin{equation}
    \frac{k_\perp}{k} = \sqrt{1 - \frac{k_\parallel^2}{k^2}} = 1 + O\left(\frac{k_\parallel^2}{k^2}\right).
\end{equation}
Therefore
\begin{equation}
    \left(\frac{k_\perp}{k}\right)^{p-2} = 1 + O\left(\frac{k_\parallel^2}{k^2}\right) = 1 + o(1),
\end{equation}
and we obtain
\begin{equation}
    d\Omega_{\hat{\mathbf{k}}} = k^{-q}\left[1 + o(1)\right] d\Omega_{p-1}\, d^q k_\parallel.
\end{equation}

\section{Gross–Pitaevskii equation}
\begin{figure}
    \centering
    \subfigure[]{\includegraphics[width=0.4\columnwidth]{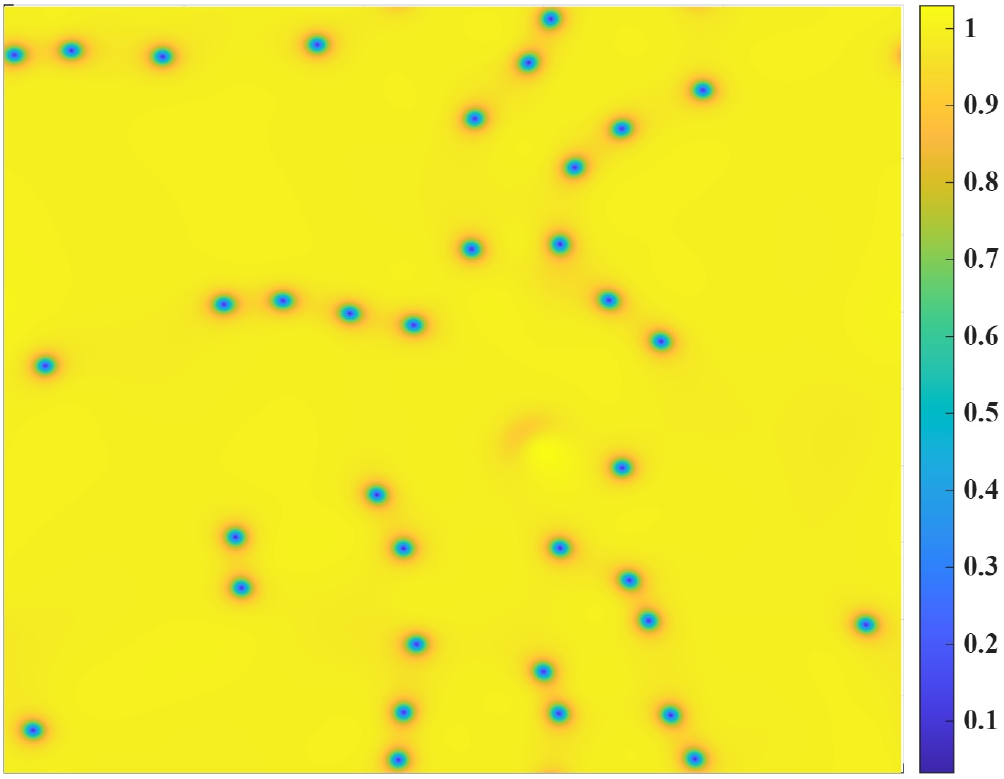}}
    \subfigure[]{\includegraphics[width=0.32\columnwidth]{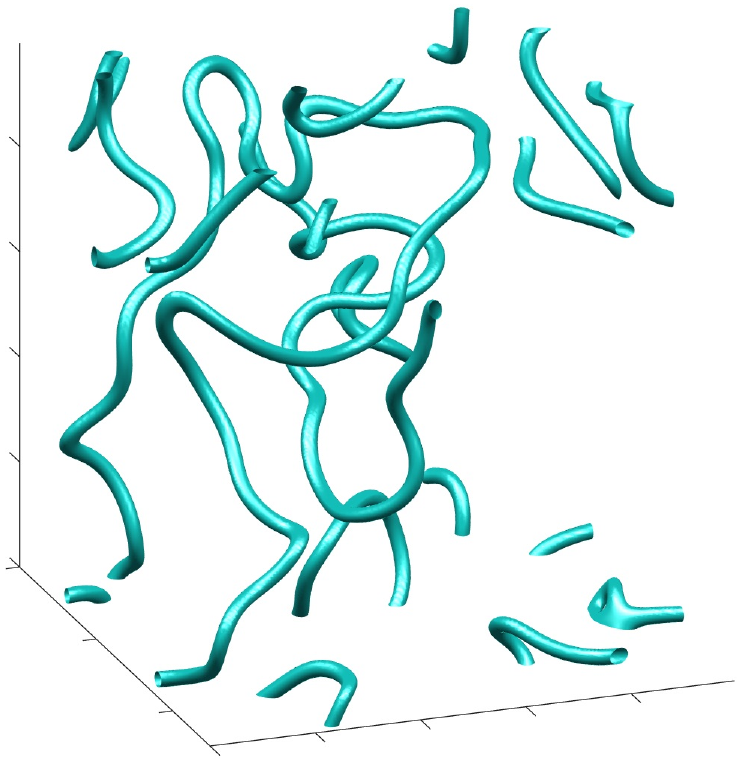}  }
    \caption{Spatial configurations of topological defects of different types and dimensions for GP model. (a) Vortex ($U(1)$ symmetry). (b) Vortex line ($U(1)$ symmetry). 
    }\label{vortexlineGP}
\end{figure}

\begin{figure}
    \centering
    \includegraphics[width=0.5\columnwidth]{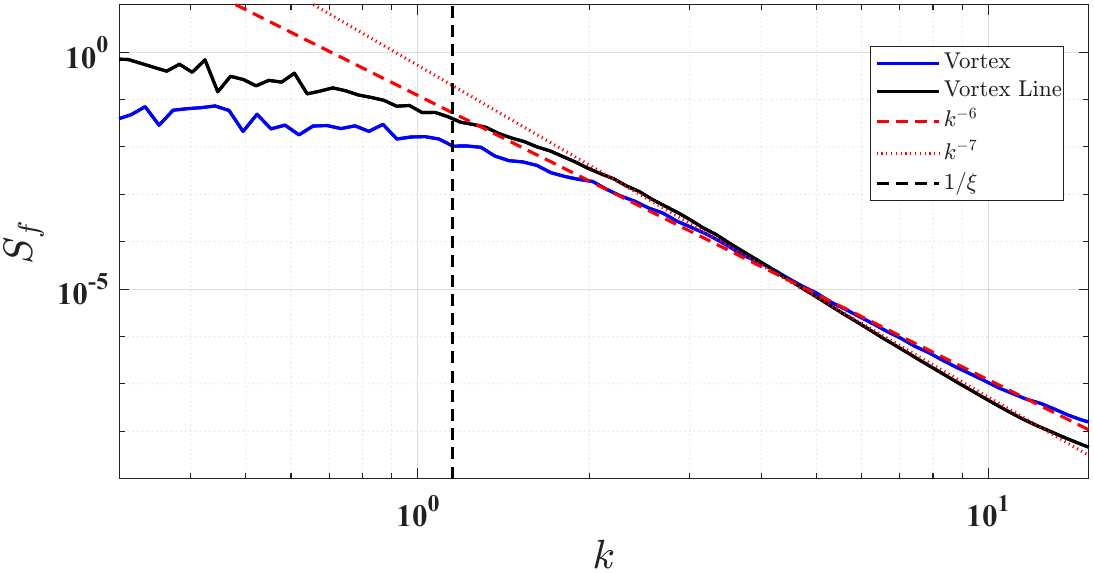}
    \caption{Scaling behavior of topological defects for GP equation of different types and dimensions.
    }\label{structurefactorexampleGP}
\end{figure}

In this section, we present the calculation results for the Gross-Pitaevskii (GP) equation. The GP equation takes the following form:
\begin{align}
(i-\Gamma)\partial_t\psi=\alpha(t)\psi+\beta\psi|\psi|^2-\gamma\nabla^2\psi.\label{qeofGl}
\end{align}
In this manuscript, we set $\Gamma = 1$. The choice of the value of $\Gamma$ is arbitrary and does not affect the scaling behavior of the system. Apart from this, the other parameters are chosen as in the TDGL model. The GP model is naturally used to describe systems with $U(1)$ symmetry. If we take the real part in this model, it becomes the TDGL model. Therefore, we only discuss the $U(1)$ case. As in the TDGL model, we numerically simulate the generation and evolution of vortices and vortex lines, and the results are displayed in Fig.~\ref{vortexlineGP}, as well as their form factors in Fig.~\ref{structurefactorexampleGP}.

\section{Holographic model}
Although the TDGL and GP models differ, both describe weak-coupling physics. It is natural to ask whether the scaling behavior of the form factor also applies to strongly coupled systems. The AdS/CFT duality provides a powerful theoretical tool for studying time-dependent evolution in such strongly coupled systems. In recent years, significant progress has been made in understanding nonequilibrium statistical physics through holographic duality \cite{Sonner:2014tca,Guo:2018mip,Yang:2023dvk,Lan:2023gyc,Zeng:2024rwn,Xia:2024wfq,Yang:2025bsw,An:2026bcw}.
Since we need to consider different symmetries, the Lagrangian of the system takes the following form:
\begin{align}
\mathcal{L}=&-e^{\alpha\Psi^2}\frac{1}{4}F_{\mu\nu}F^{\mu\nu}
-D_{\mu}\Psi^{\ast} D^{\mu}\Psi
-m^{2}\Psi^*\Psi,
\end{align}
in which $D_{\mu}\Psi=\nabla_{\mu}\Psi-inA_\mu\Psi$ is the covariant derivative term of the charged scalar field, and $F_{\mu\nu}=\nabla_{\mu}A_{\nu}-\nabla_{\nu}A_{\mu}$ is the Maxwell field strength. When $\alpha = 0$ and $n = 1$, the system corresponds to $U(1)$ symmetry. When $n = 0$ and $\alpha \neq 0$, the system corresponds to a neutral scalar field with $\mathbb{Z}_2$ symmetry. In the following calculations, without loss of generality, we set $\alpha = 5$. In the holographic model, the temperature of the system is given by $T = 3/(4\pi z_h)$, where $z_h$ is the radius of the black hole horizon. When the system is below the critical temperature $T_c$, spontaneous symmetry breaking occurs, and the physical process is the same as in the TDGL and GP models. In this manuscript, we neglect the backreaction of the matter fields on the metric, which means we are working in the probe limit. 

\begin{figure}
    \centering
        \subfigure[]{\includegraphics[width=0.4\columnwidth]{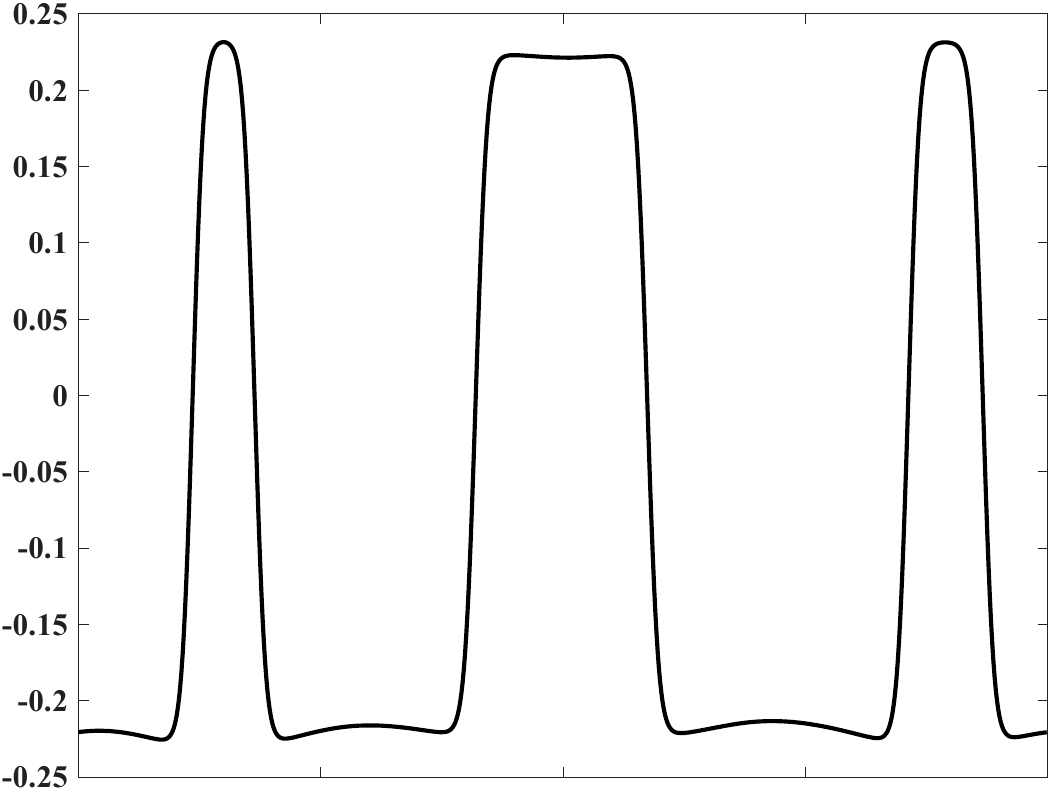}}
        \subfigure[]{\includegraphics[width=0.4\columnwidth]{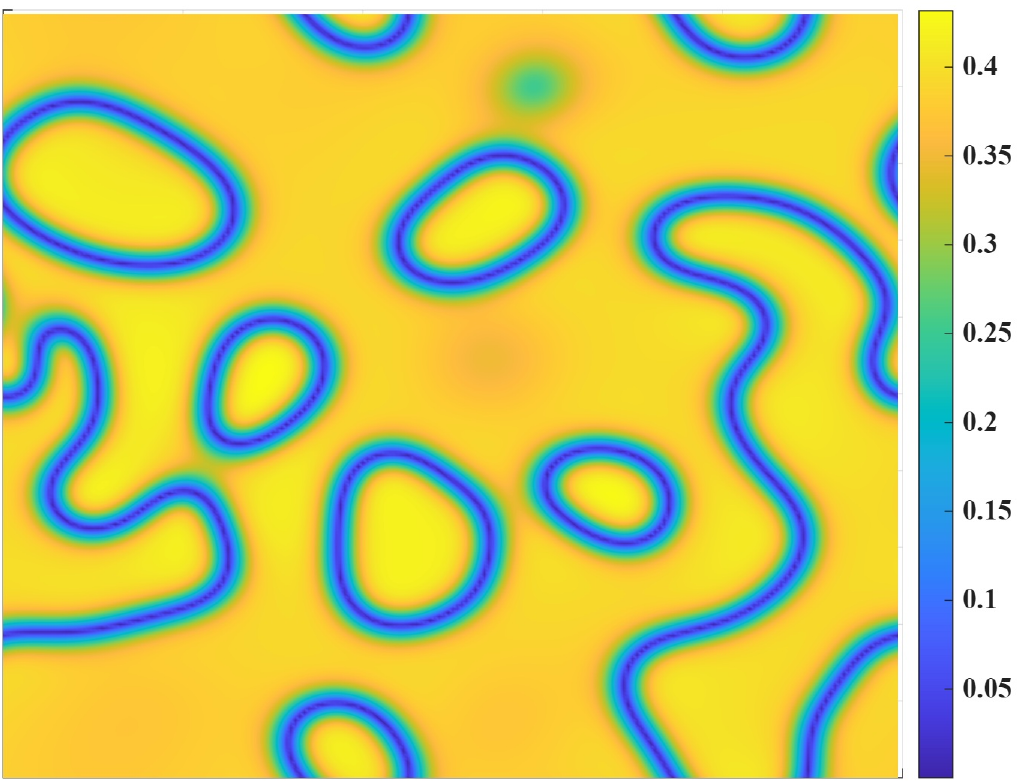}  }
        \subfigure[]{\includegraphics[width=0.4\columnwidth]{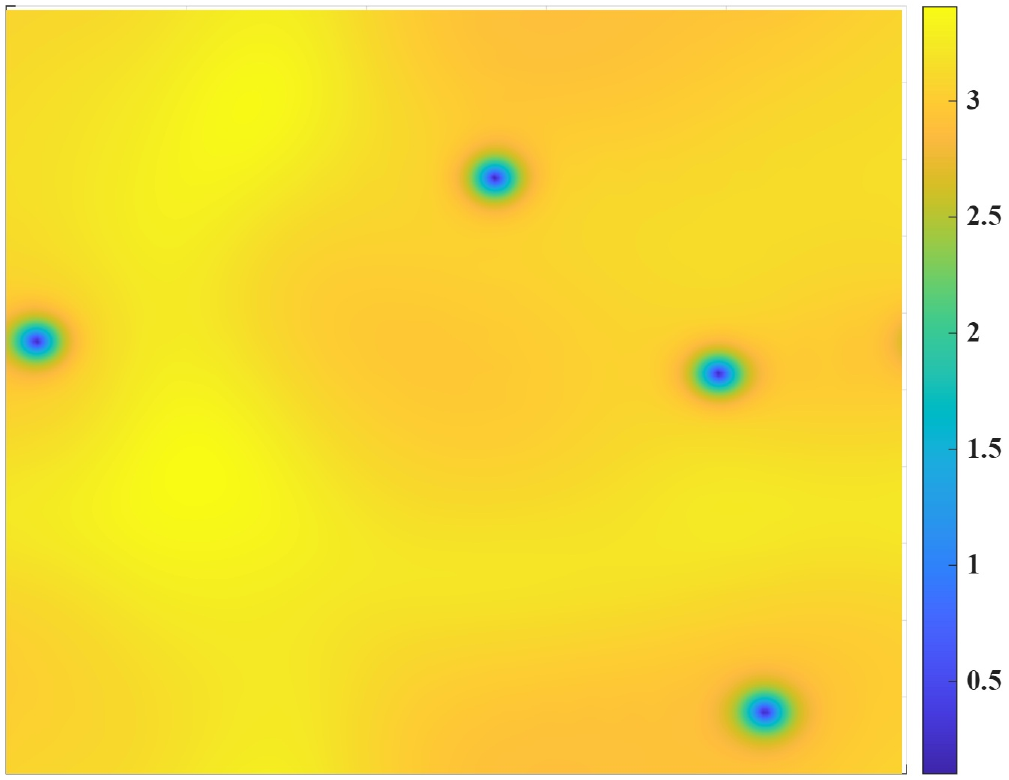} } 
        \subfigure[]{\includegraphics[width=0.33\columnwidth]{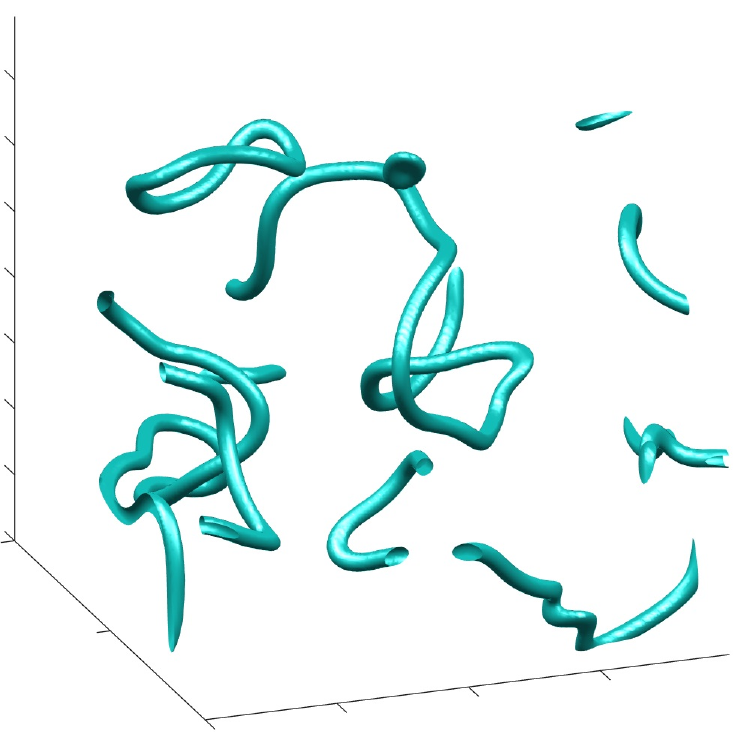} }
    \caption{Spatial configurations of topological defects of different types and dimensions for holographic model. (a) Kink ($\mathbb{Z}_2$ symmetry). (b) Domain wall ($\mathbb{Z}_2$ symmetry). (c) Vortex ($U(1)$ symmetry). (d) Vortex line ($U(1)$ symmetry). }\label{topologicaldefectexampleholographic}
\end{figure}

In this paper, the holographic model under consideration involves dynamical evolution. Therefore, the most convenient choice is to adopt the ingoing Eddington metric
\begin{align}
ds^{2}=\frac{1}{z^2}\left(-f(z)dt^{2}-2dtdz+dx^{2}+dy^{2}\right),
\end{align}
with $f(z)=1-(z/z_h)^3$. We adopt the following ansatz
\begin{align}
&\Psi =z\psi(t,z,\vec{x}),\\
&A_\mu dx^\mu=
A_t(t,z,\vec{x})dt+A_{\vec{x}}(t,z,\vec{x})d\vec{x}.
\end{align}
The complete equations of motion of $U(1)$ symmetry are given as follows
\begin{align}
-2 i A_t \partial _z\psi +A_x^2 \psi +2 i A_x \partial _x\psi +A_y ^2 \psi +2 i A_y \partial _y\psi +2 \partial _t\partial _z\psi +i \partial_xA_x \psi -\partial _x\partial _x\psi +i \partial _yA_y \psi\nonumber\\
-\partial _y\partial _y\psi-i \partial _zA_t \psi +\partial _z\partial _z\psi z^3-\partial _z\partial _z\psi +3 \partial _z\psi  z^2+\psi  z=0,&\label{A1C}\\
\partial _z\partial _xA_x+\partial _z\partial _yA_y-\partial _z\partial _zA_t-i \partial _z\psi ^* \psi +i \partial _z\psi  \psi ^*=0,&\label{A2C}\\
2 A_x \psi ^* \psi +2 \partial _t\partial _zA_x+\partial _x\partial _yA_y-i \partial _x\psi ^* \psi +i \partial _x\psi  \psi ^*-\partial_y\partial _yA_x+3 \partial _zA_x z^2-\partial _z\partial _xA_t\nonumber\\
+\partial _z\partial _zA_x z^3-\partial _z\partial _zA_x=0,&\label{A3C}\\  
2 A_y \psi ^* \psi +2 \partial _t\partial _zA_y+\partial _x\partial _yA_x-i \partial _y\psi ^* \psi +i \partial _y\psi 
\psi ^*-\partial _x\partial _xA_y+3 \partial _zA_y z^2-\partial _z\partial _yA_t\nonumber\\
+\partial _z\partial _zA_y z^3-\partial _z\partial _zA_y=0,&\label{A4C}\\
-2 A_t \psi ^* \psi -\partial _t\partial _xA_x-\partial _t\partial _yA_y-\partial _t\partial _zA_t+i \partial _t\psi ^* \psi -i \partial _t\psi \psi ^*+\partial _x\partial _xA_t+\partial _y\partial _yA_t\nonumber\\
-\partial _z\partial _xA_x z^3+\partial _z\partial _xA_x-\partial _z\partial _yA_yz^3+\partial _z\partial _yA_y-i \partial _z\psi ^* \psi +i \partial _z\psi ^* \psi  z^3\nonumber\\
-i \partial _z\psi  \psi ^* z^3+i \partial _z\psi  \psi
^*=0.&\label{A5C}
\end{align}
For the system with $\mathbb{Z}_2$ symmetry, the formulas are as follows
\begin{align}
\frac{\partial _z\psi  f'}{2}+\frac{\psi  f'}{2 z}-\frac{1}{4} \alpha  \psi  z^2 e^{\alpha  \psi ^2 z^2} \big(f
   \partial _zA_x{}^2+f \partial _zA_y{}^2-2 \partial _tA_x \partial _zA_x-2 \partial _tA_y \partial _zA_y+2 \partial
   _xA_t \partial _zA_x\nonumber\\+\partial _xA_y{}^2-2 \partial _xA_y \partial _yA_x
   +2 \partial _yA_t \partial _zA_y+\partial
   _yA_x{}^2-\partial _zA_t{}^2\big)
   +\frac{f \partial _z\partial _z\psi }{2}-\frac{f \psi }{z^2}\nonumber\\
   -\partial _t\partial _z\psi +\frac{\partial _x\partial _x\psi }{2}+\frac{\partial _y\partial _y\psi
   }{2}+\frac{\psi }{z^2}=0,\label{A1}\\
   2 \alpha  \partial _x\psi  \partial _zA_x \psi  z^2+2 \alpha  \partial _y\psi  \partial _zA_y \psi  z^2-2 \alpha 
   \partial _zA_t \partial _z\psi  \psi  z^2-2 \alpha  \partial _zA_t \psi ^2 z+\partial _z\partial _xA_x\nonumber\\
   +\partial_z\partial _yA_y-\partial _z\partial _zA_t=0,\label{A2}\\
   \frac{\partial _zA_x f'}{2}+\alpha  f \partial _zA_x \partial _z\psi  \psi  z^2+\alpha  f \partial _zA_x \psi ^2
   z+\frac{f \partial _z\partial _zA_x}{2}-\alpha  \partial _tA_x \partial _z\psi  \psi  z^2+\alpha  (-\partial
   _tA_x) \psi ^2 z\nonumber\\
   -\partial _t\partial _zA_x-\alpha  \partial _t\psi  \partial_zA_x \psi  z^2+\alpha  \partial
   _xA_t \partial _z\psi  \psi  z^2+\alpha  \partial _xA_t \psi ^2 z-\alpha  \partial _xA_y \partial _y\psi  \psi 
   z^2\nonumber\\
   -\frac{\partial _x\partial _yA_y}{2}+\alpha  \partial _yA_x \partial _y\psi  \psi  z^2+\frac{\partial
   _y\partial _yA_x}{2}+\frac{\partial _z\partial _xA_t}{2}=0,\label{A3}\\
   \frac{\partial _zA_y f'}{2}+\alpha  f \partial _zA_y \partial _z\psi  \psi  z^2+\alpha  f \partial _zA_y \psi ^2
   z+\frac{f \partial _z\partial _zA_y}{2}-\alpha  \partial _tA_y \partial _z\psi  \psi  z^2+\alpha  (-\partial
   _tA_y) \psi ^2 z\nonumber\\
   -\partial _t\partial _zA_y-\alpha  \partial _t\psi  \partial _zA_y \psi  z^2+\alpha  \partial_xA_y \partial _x\psi  \psi  z^2+\frac{\partial _x\partial _xA_y}{2}-\frac{\partial _x\partial _yA_x}{2}-\alpha \partial _x\psi  \partial _yA_x \psi  z^2\nonumber\\
   +\alpha  \partial _yA_t \partial _z\psi  \psi  z^2+\alpha  \partial _yA_t\psi ^2 z+\frac{\partial _z\partial _yA_t}{2}=0,\label{A4}\\
   2 \alpha  f \partial _x\psi  \partial _zA_x \psi  z^2+2 \alpha  f \partial _y\psi  \partial _zA_y \psi  z^2+f \partial _z\partial _xA_x+f \partial _z\partial _yA_y-2 \alpha  \partial _tA_x \partial _x\psi  \psi  z^2\nonumber\\
   -2 \alpha  \partial _tA_y \partial_y\psi  \psi  z^2-\partial _t\partial _xA_x-\partial _t\partial _yA_y-\partial _t\partial _zA_t-2 \alpha  \partial _t\psi  \partial _zA_t \psi  z^2\nonumber\\
   +2 \alpha  \partial _xA_t \partial _x\psi  \psi  z^2+\partial _x\partial _xA_t+2 \alpha  \partial _yA_t\partial _y\psi  \psi  z^2+\partial _y\partial _yA_t=0.\label{A5}
\end{align}
To solve the field equations, boundary conditions need to be specified. At the AdS boundary, the asymptotic expansions of $A_t$  and $\psi$ are given by $A_t = \mu - z\rho$ and $\psi = \psi^{(1)} + z\psi^{(2)} + \cdots$, where $\mu$ and $\rho$ denote the chemical potential and charge density, respectively. Eq.~(\ref{A5}) and Eq.~(\ref{A5C}) are constraint equations
\begin{align}
\partial _t\rho=-\partial _x\partial _xA_t-\partial _y\partial _yA_t-\partial _z\partial _xA_x-\partial _z\partial _yA_y,
\end{align}
which implies the current conservation condition for $A_t$. For the scalar field $\psi$, we impose $\psi^{(1)} = 0$. For $A_x$ and $A_y$, Dirichlet boundary conditions are applied. In the $x-y$ plane, periodic boundary conditions are adopted due to the Fourier spectral method. In the $z$ direction, we use the Chebyshev pseudospectral method. For the time evolution, the fourth-order Runge-Kutta method is employed with a time step of $\delta t=0.05$. In addition, we set $m^2 = -2$ in our calculations.

\begin{figure}[!htbp]
    \centering
    \includegraphics[width=0.5\columnwidth]{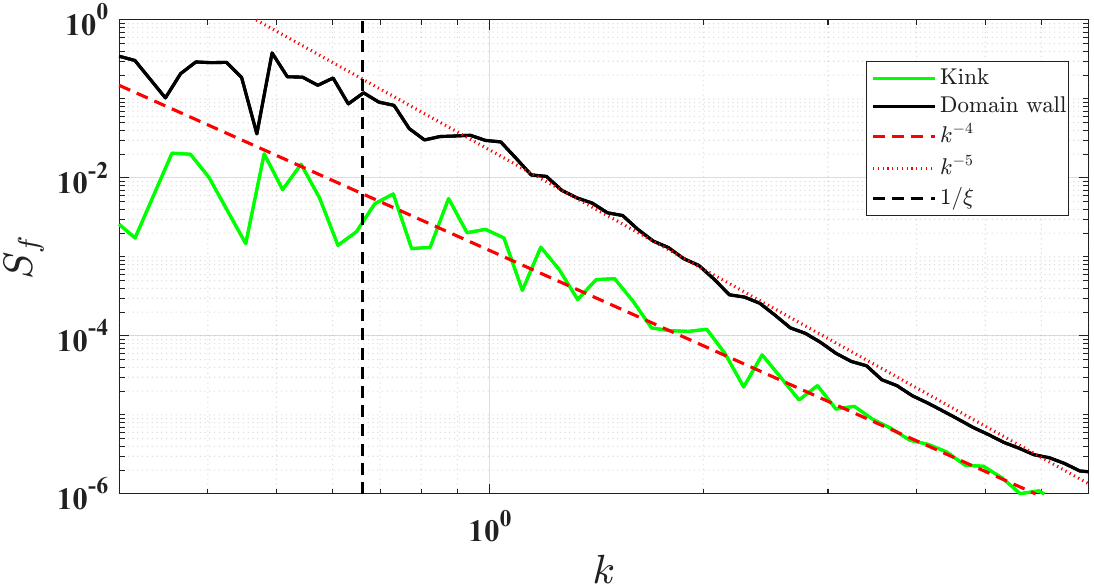}
    \includegraphics[width=0.5\columnwidth]{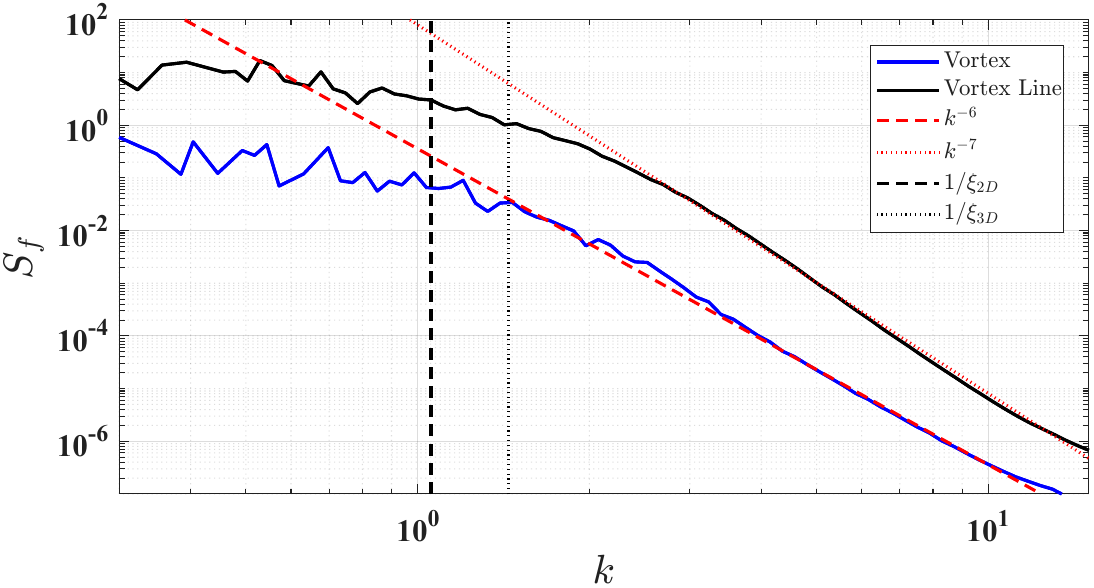}
    \caption{Scaling behavior of topological defects for holographic model of different types and dimensions. The upper panel corresponds to the system with $\mathbb{Z}_2$ symmetry, and the lower panel corresponds to the system with $U(1)$ symmetry.
    }\label{structurefactorexampleholographic}
\end{figure}

In the holographic model, the asymptotic expansion at the AdS boundary is given by $\psi = \psi^{(1)} + z\psi^{(2)}$ and $A_t = \mu - z\rho$, where $\mu$ and $\rho$ are the chemical potential and the charge density, respectively. In the subsequent numerical calculations, we fix the total charge density $\rho$. Moreover, according to the scaling symmetry, the temperature can also be written as $T = 3/(4\pi \rho^{1/2})$. The critical value is $\rho_c = 4.07$, which gives the critical temperature $T_c = 0.118$. We also adopt the source-free boundary condition, namely we fix $\psi^{(1)} = 0$.

Similar to the TDGL case, we compute topological defects of different dimensions for both $\mathbb{Z}_2$ and $U(1)$ symmetries in the holographic model, and calculate their form factors. In this paper, we do not compute the 3D domain wall configurations with $\mathbb{Z}_2$ symmetry in the holographic model, as they are too complicated and computationally expensive. We believe that the results for 3D domain wall configurations in the holographic model should be the same. In addition, we do not perform a full simulation of the dynamical evolution of 3D vortex lines in the holographic model; instead, we use the data for vortex lines from Ref.~\cite{Zeng:2024rwn}. The spatial distribution of the order parameter obtained from numerical calculations is shown in Fig.~\ref{topologicaldefectexampleholographic}. The results for the form factors are presented in Fig.~\ref{structurefactorexampleholographic}.

\section{Klein-Gordon equation in FRW background}
\begin{figure}
    \centering
    \includegraphics[width=0.25\columnwidth]{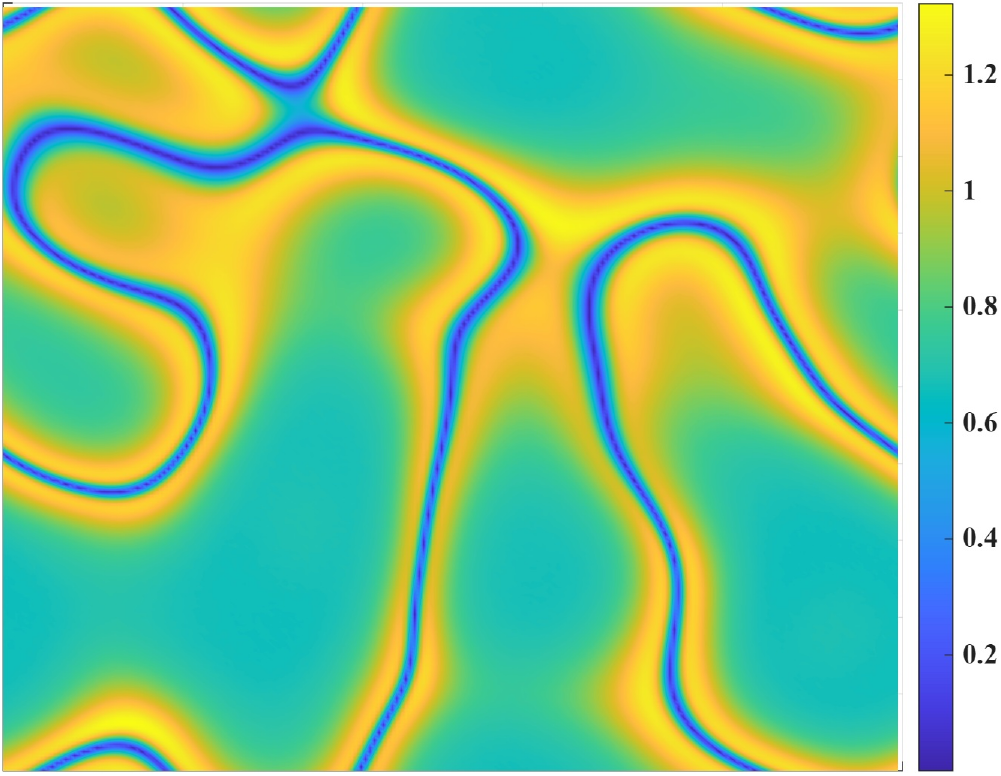}
    \includegraphics[width=0.25\columnwidth]{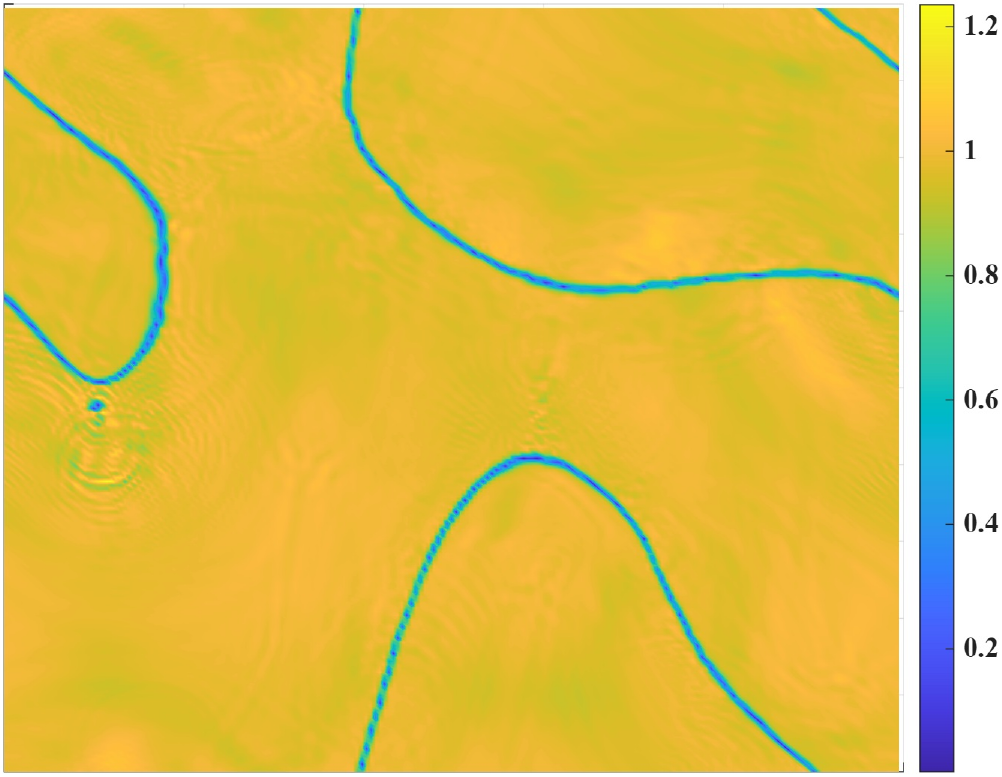}  
    \includegraphics[width=0.25\columnwidth]{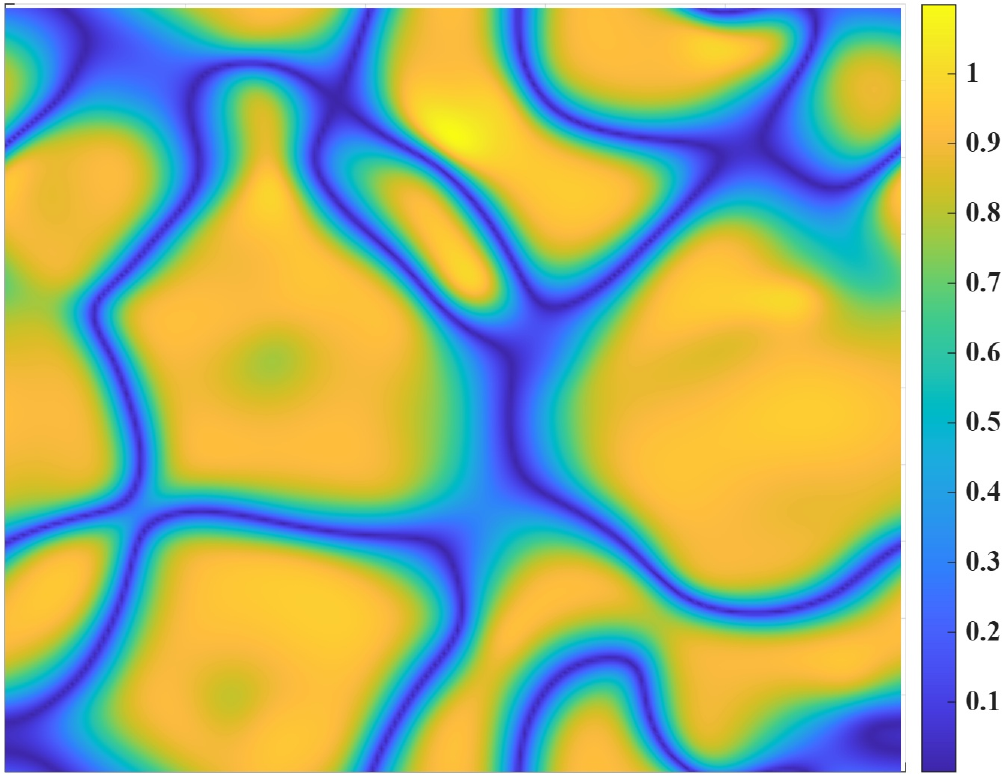}  
    \includegraphics[width=0.25\columnwidth]{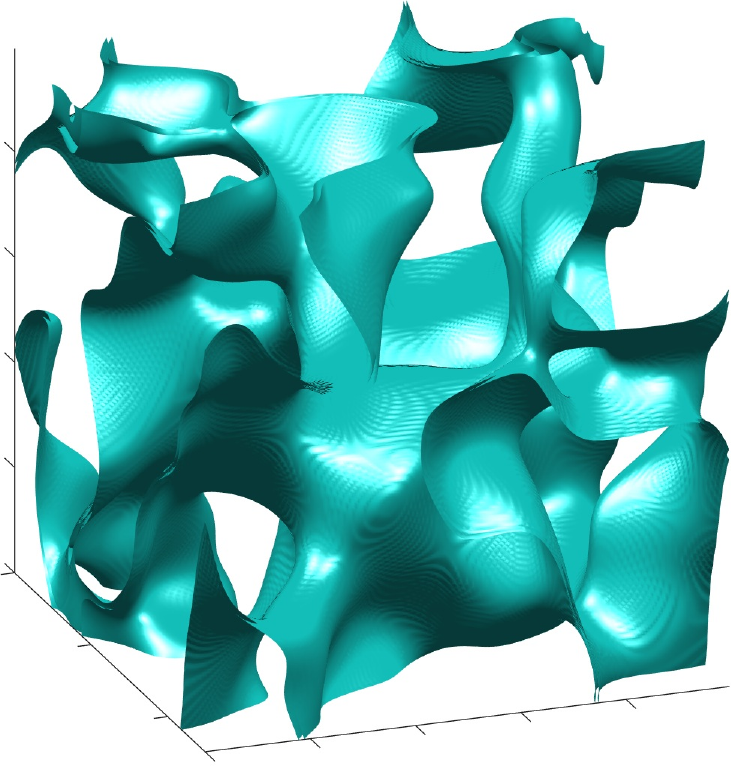} 
    \includegraphics[width=0.25\columnwidth]{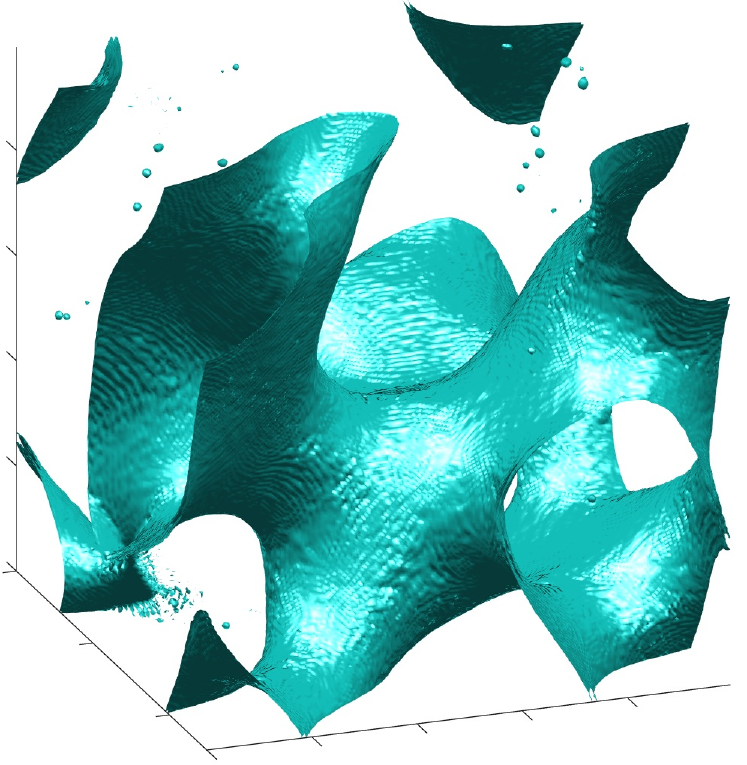} 
    \includegraphics[width=0.25\columnwidth]{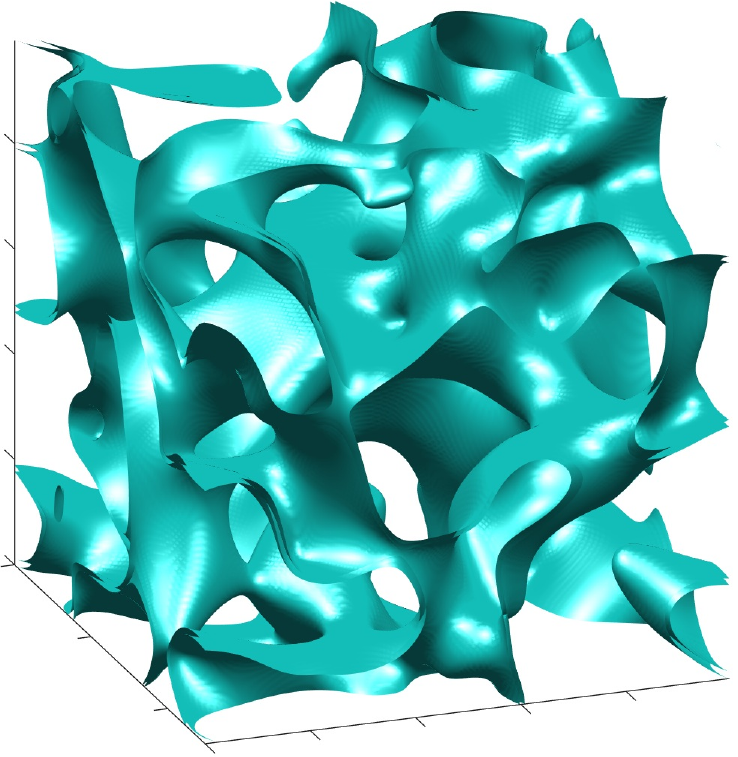} 
    \caption{Spatial configurations of topological defects of different types and dimensions for FRW model.
    The left panel shows the spatial distribution of $\phi$ for the FRW case at $\eta=15.5$, the middle panel shows that at $\eta=30$, and the right panel shows that for the PRS case at $\eta=15.5$. The upper row shows the results for the 2D case, and the lower row shows the results for the 3D case.}\label{topologicaldefectexampleFRW}
\end{figure}

\begin{figure}
    \centering
    \includegraphics[width=0.5\columnwidth]{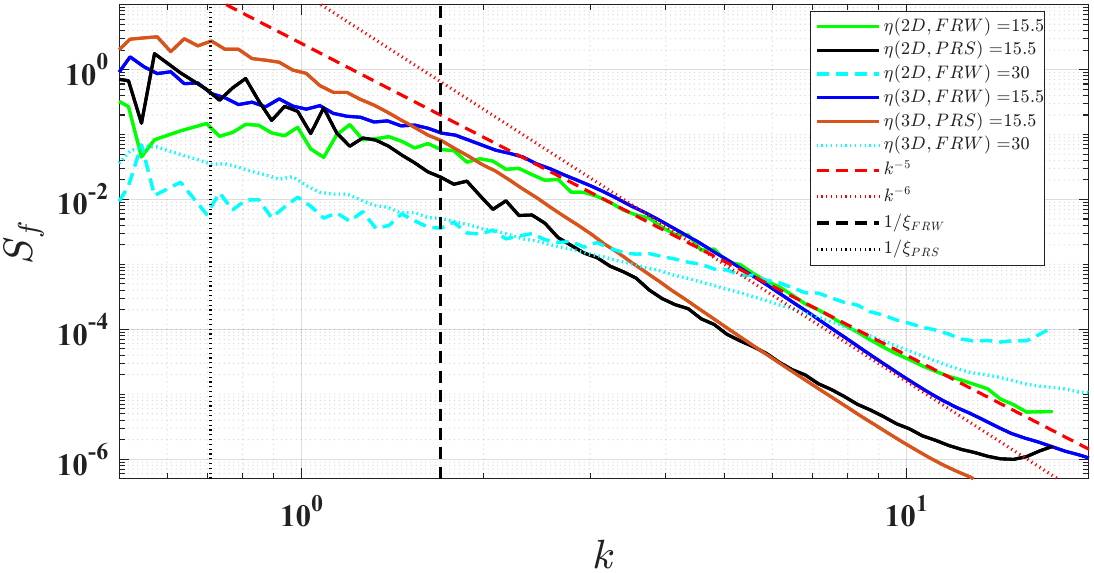}
    \caption{Scaling behavior of topological defects for FRW equation in 2-dimension.
    }\label{SFFRW}
\end{figure}

In this section, we present the Klein-Gordon (KG) equation in the FRW background. It takes the following form: \cite{Press:1989yh}:
\begin{align}
    \phi''+\alpha\frac{a'}{a}\phi'-\nabla^2\phi=-a^{\beta}\frac{\partial V}{\partial\phi},
\end{align}
in which $V=1/4(\phi^2-\phi^2_0)^2$ and $a\propto (\eta/\eta_0)^2$  with $\eta_0=10$. In the following calculations, we uniformly use the conformal time $\eta$.
For the case of symmetry breaking, the KG equations have a standard planar solution of the form $\phi = \phi_0 \tanh(\phi_0/\sqrt{2} \, a^{\beta/2}(z-z_0))$, whose form is consistent with the standard domain wall configuration Eq.~(\ref{eqhealingfit}). However, unlike the TDGL and holographic cases, the KG equations incorporate the effect of cosmic expansion. 
This means that the domain wall thickness, i.e., the comoving healing window $\xi_{\rm com} =\sqrt{2}/\phi_0~a^{-\beta/2}$, will be infinitely squeezed, eventually making the original V-shaped cusp structure of the domain wall unrecognizable at the given spatial resolution. However, in principle, sufficient resolution could resolve it, but at the cost of substantial computational resources and time. Therefore, for the current numerical resolution, the scaling behavior of the two-dimensional and three-dimensional domain wall configuration with $\alpha=2$ and $\beta=2$ exists only at early times when the comoving time $\eta$ is sufficiently small. Here ``two-dimensional'' refers to a slice in three-dimensional space, specifically the $xy$-plane at $z=0$. However, we have also computed the purely two-dimensional case, and the results are the same.
Two-dimensional means that our universe is still three-dimensional, but due to considerations of homogeneity and isotropy, our equations can be made independent of the $z$ direction, as in Ref.~\cite{Press:1989yh}. Nevertheless, there is another way to preserve the domain wall configuration from being squeezed, namely the PRS modification \cite{Press:1989yh}, which takes $(\alpha=3, \beta=0)$. The numerical results are shown in Fig.~\ref{topologicaldefectexampleFRW}. We compute the form factors for both cases and present the results in Fig.~\ref{SFFRW}. The results show that the scaling behavior of $S_f$ always exists, provided that the resolution and the configuration are well defined. 

\section{Fitting test}
We present the results of the sliding-window test in the parameter space for fitting the scaling behavior. In the sliding-window plot, the blue asterisks denote the fitting intervals selected for the scaling behavior in the table of the main text.

\begin{figure}
    \centering
    \includegraphics[width=0.45\columnwidth]{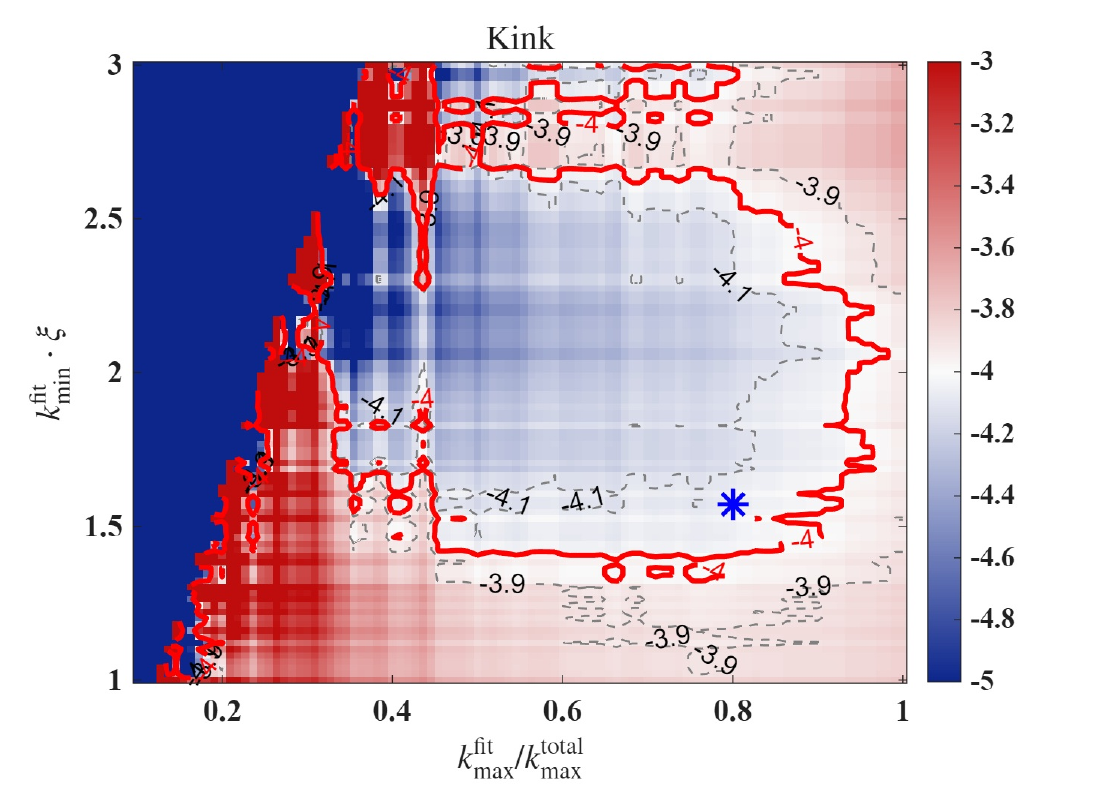}
    \includegraphics[width=0.45\columnwidth]{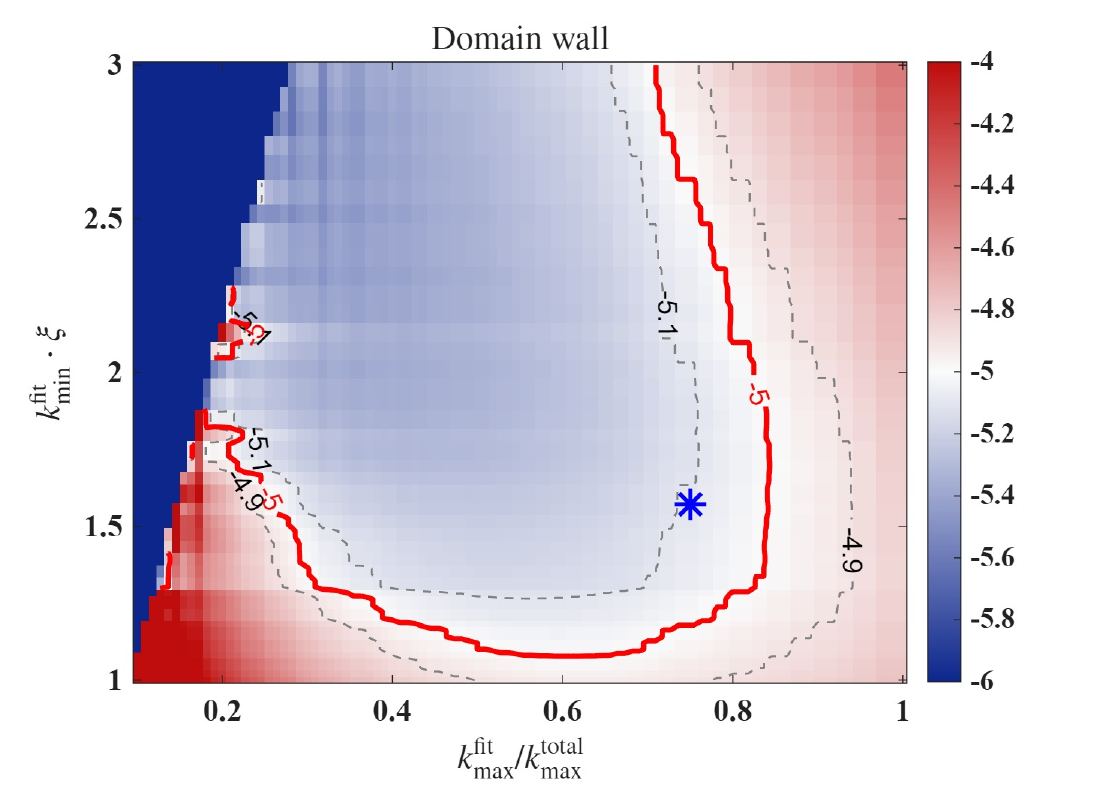}
    \includegraphics[width=0.45\columnwidth]{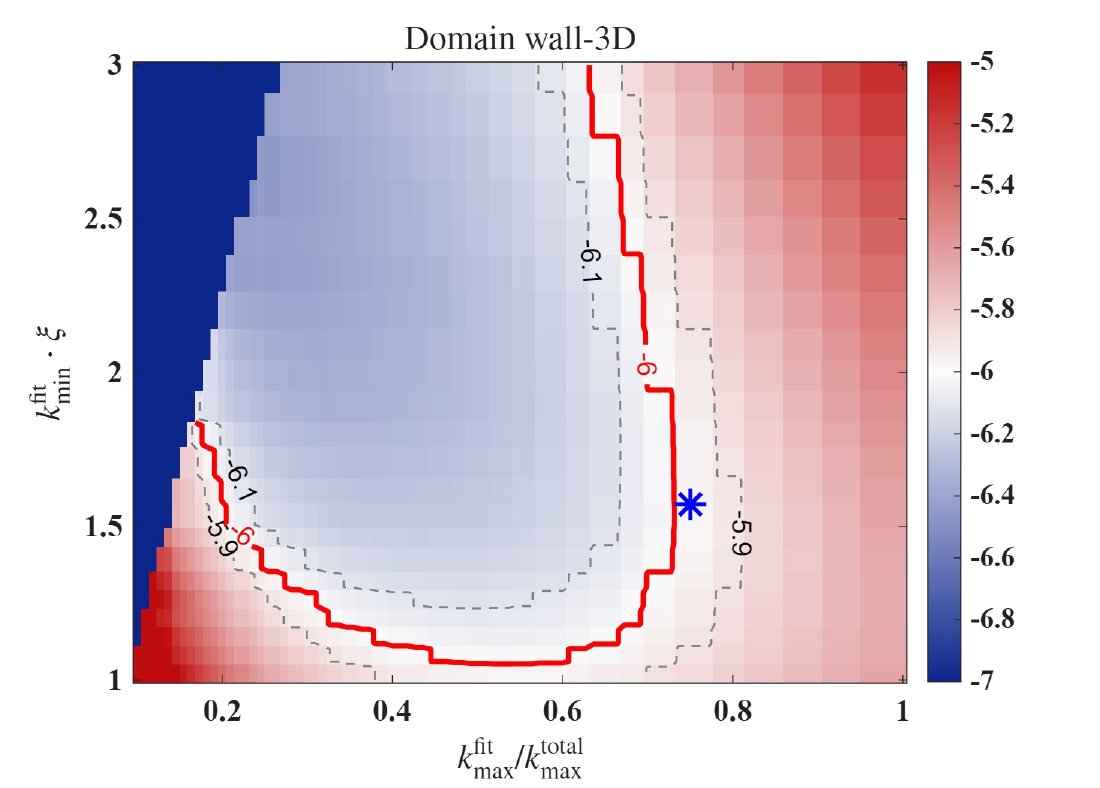}
    \includegraphics[width=0.45\columnwidth]{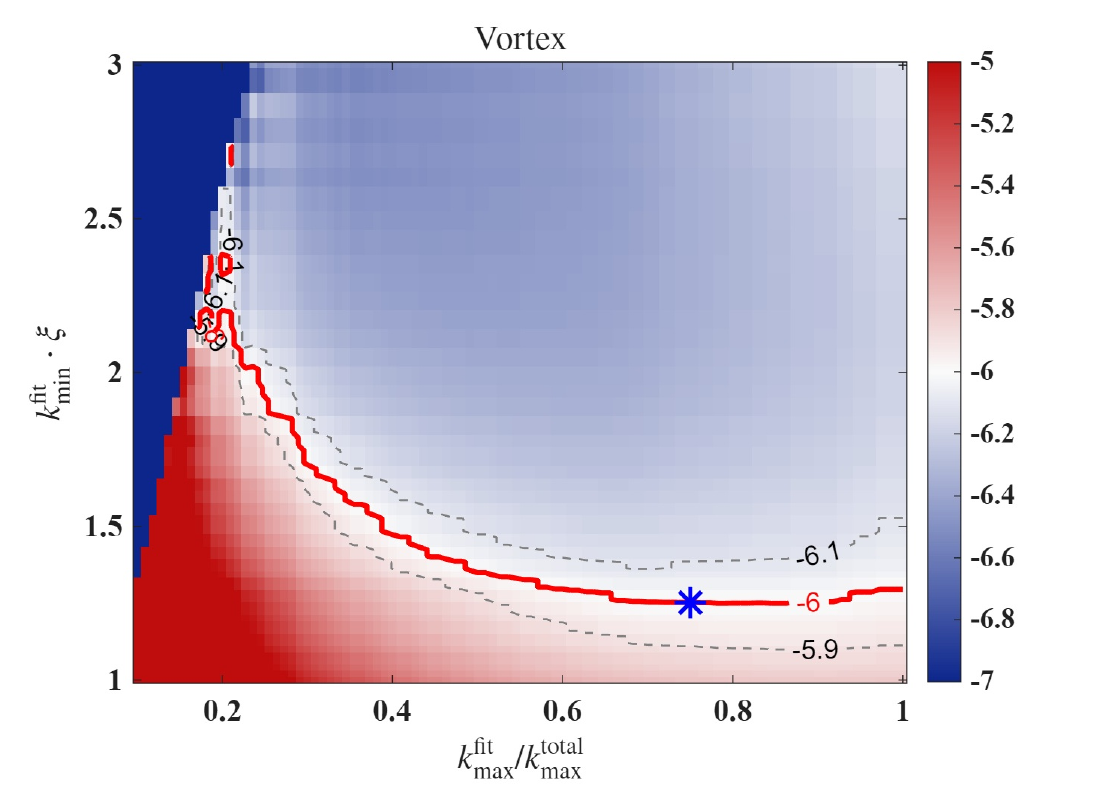}
    \includegraphics[width=0.45\columnwidth]{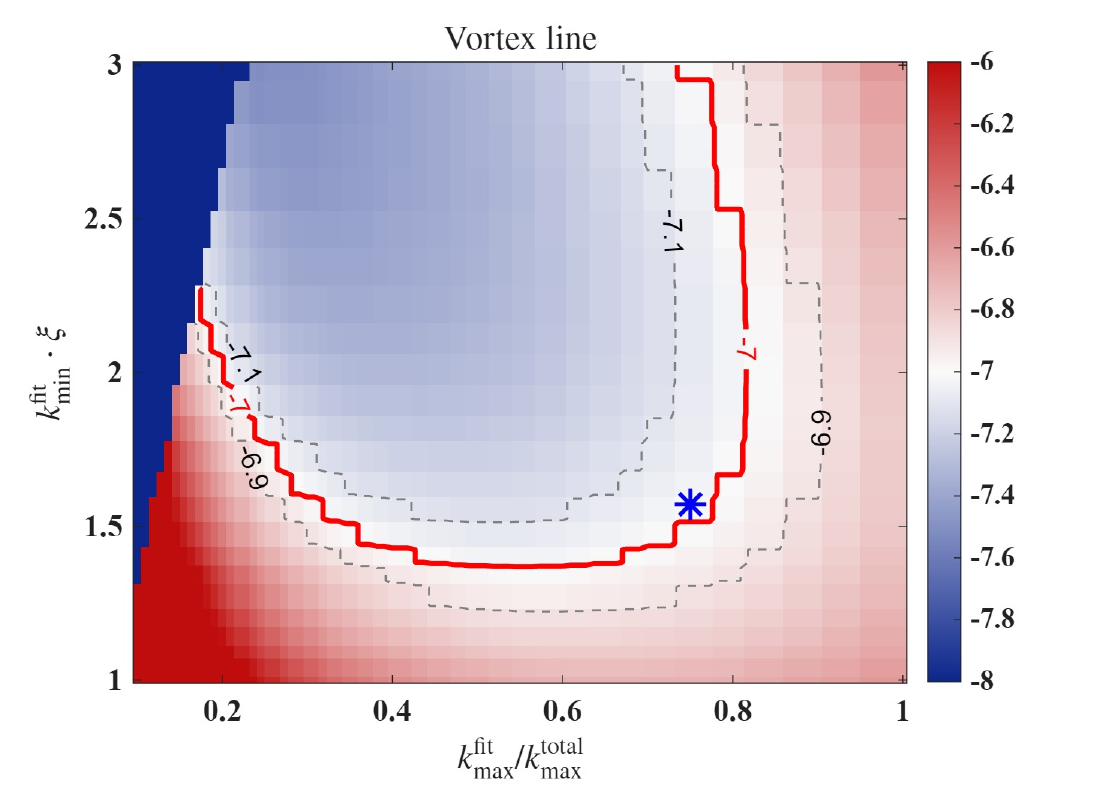}
    \includegraphics[width=0.45\columnwidth]{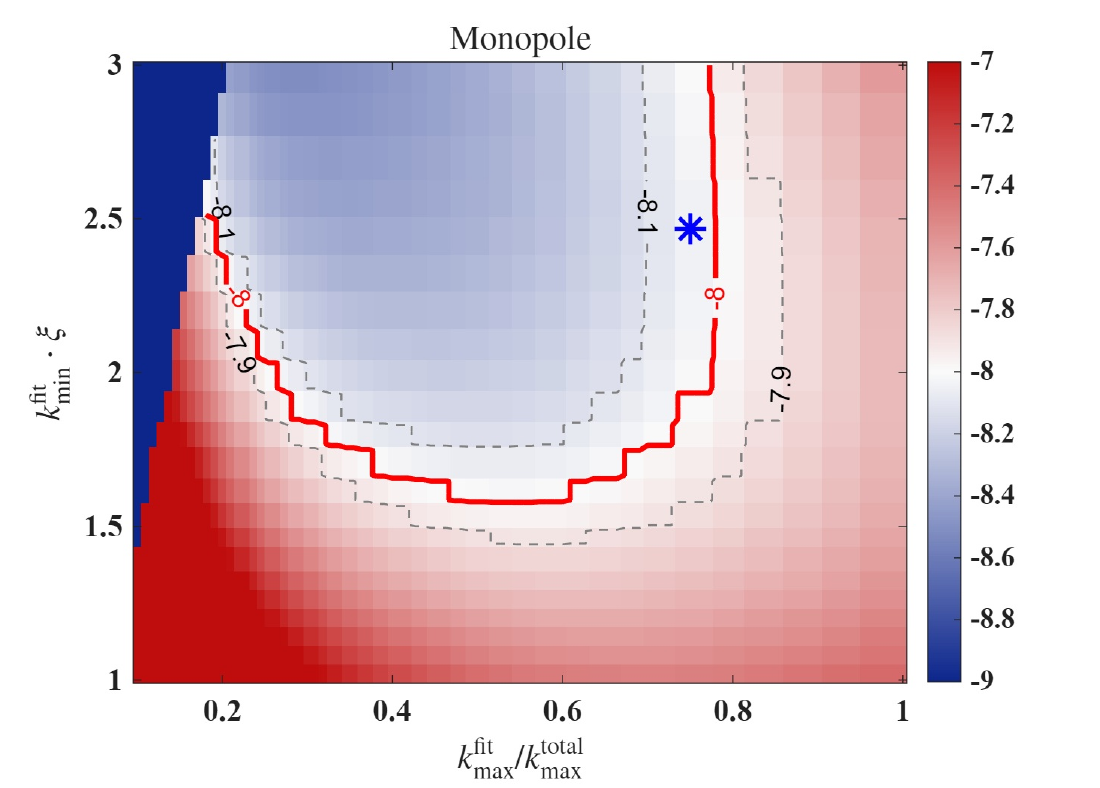}
    \caption{Analysis of the fitting interval for GL model.
    }\label{fittingHitU1GL}
\end{figure}

\begin{figure}
    \centering
    \includegraphics[width=0.45\columnwidth]{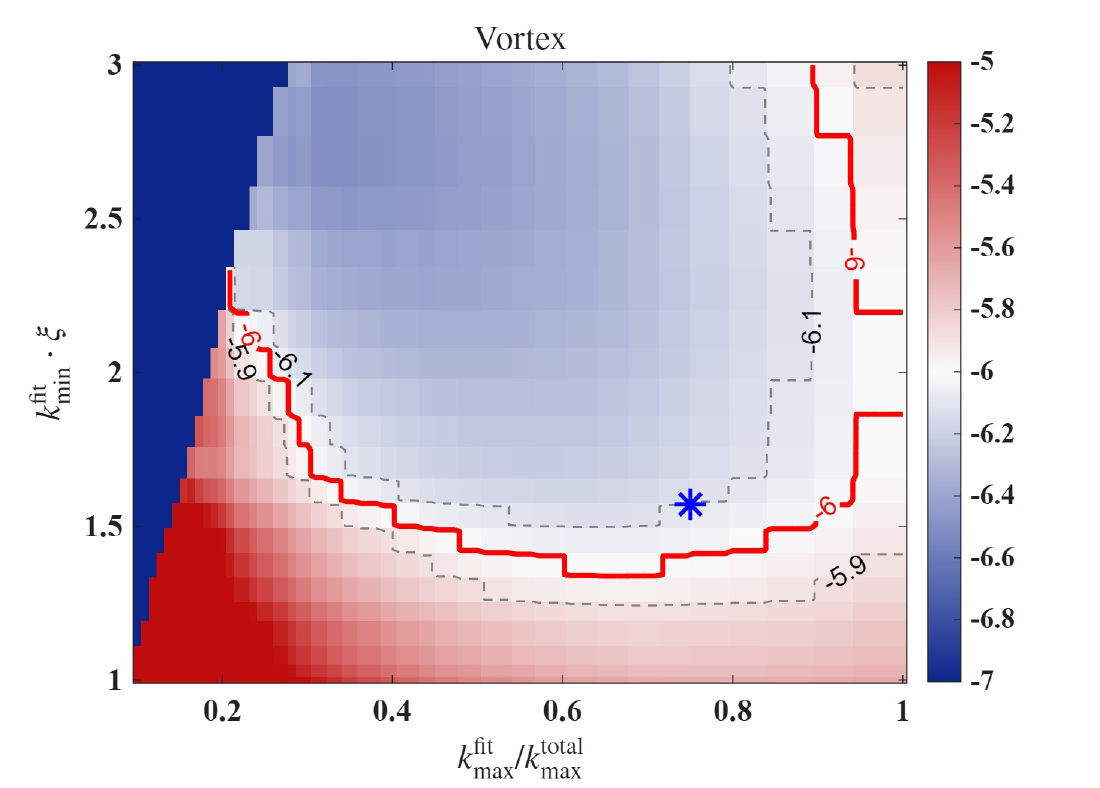}
    \includegraphics[width=0.45\columnwidth]{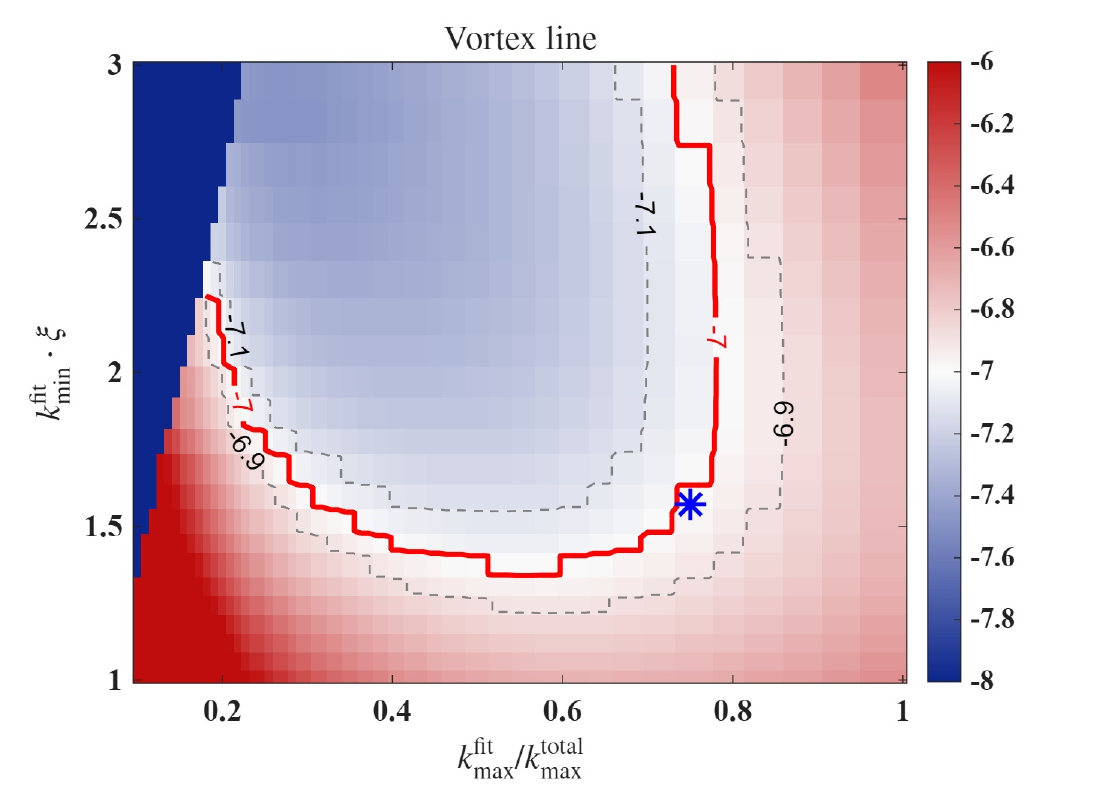}
    \caption{Analysis of the fitting interval for the $U(1)$ symmetry for GP model.
    }\label{fittingHitU1GP}
\end{figure}

\begin{figure}
    \centering
    \includegraphics[width=0.45\columnwidth]{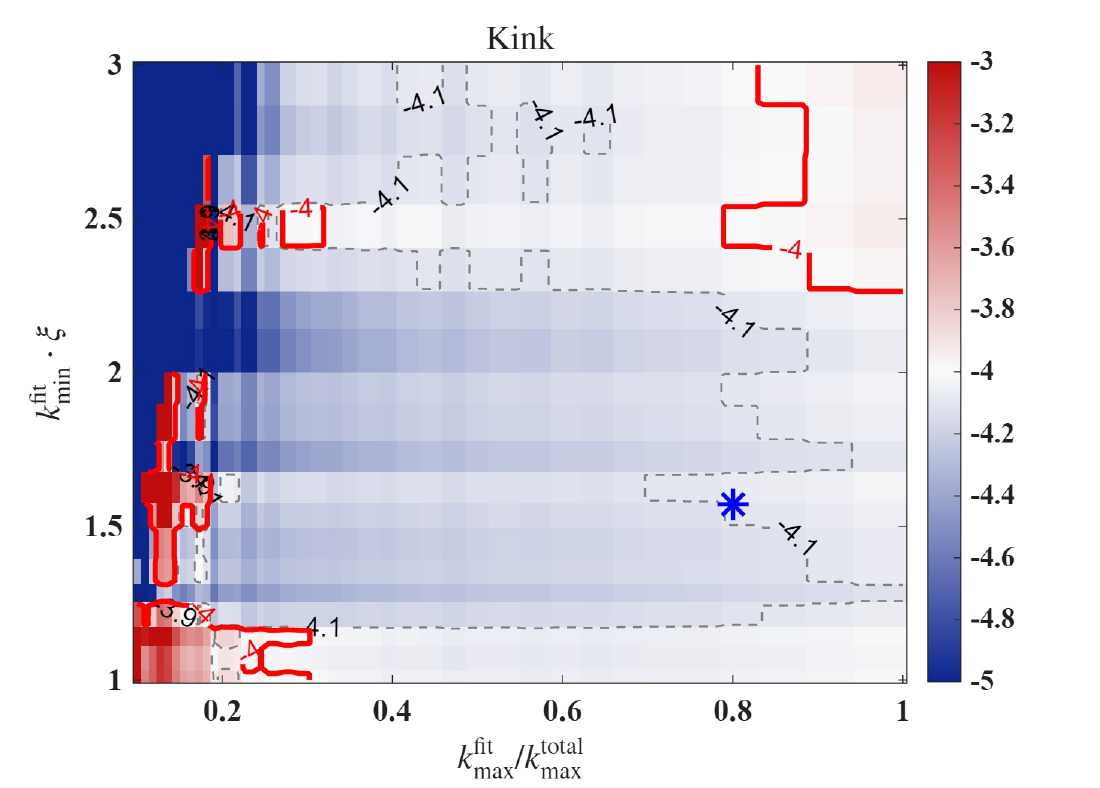}
    \includegraphics[width=0.45\columnwidth]{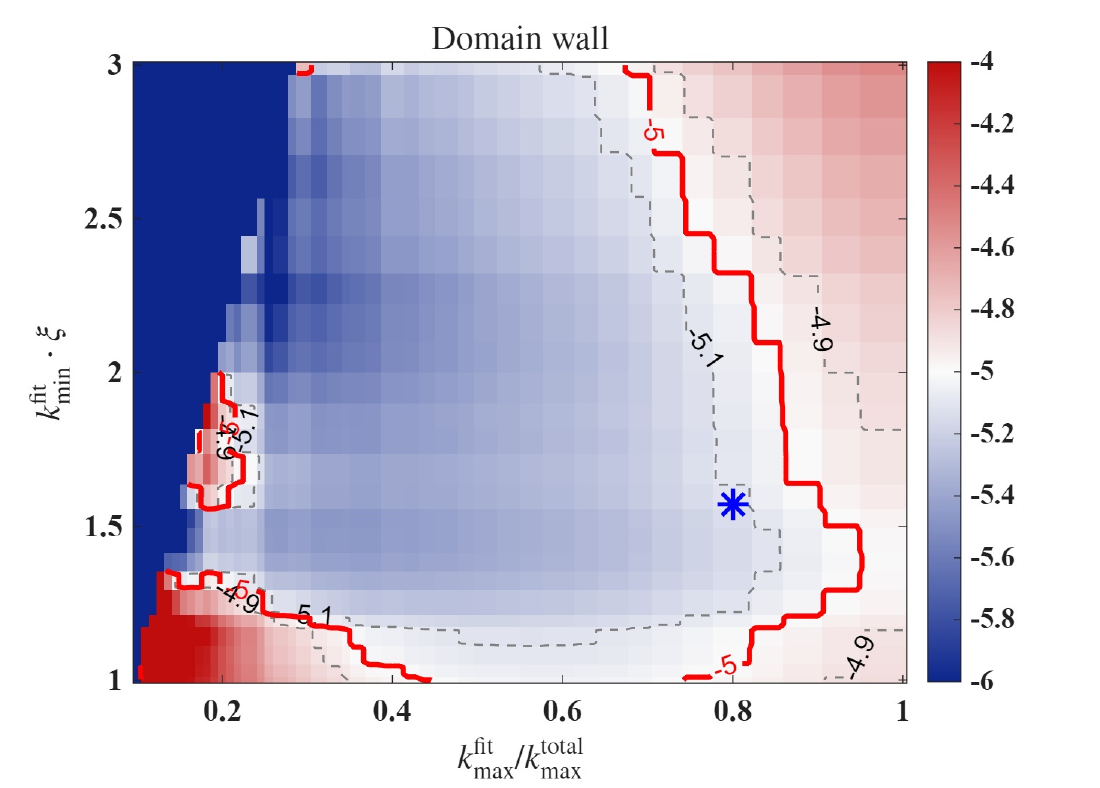}
    \includegraphics[width=0.45\columnwidth]{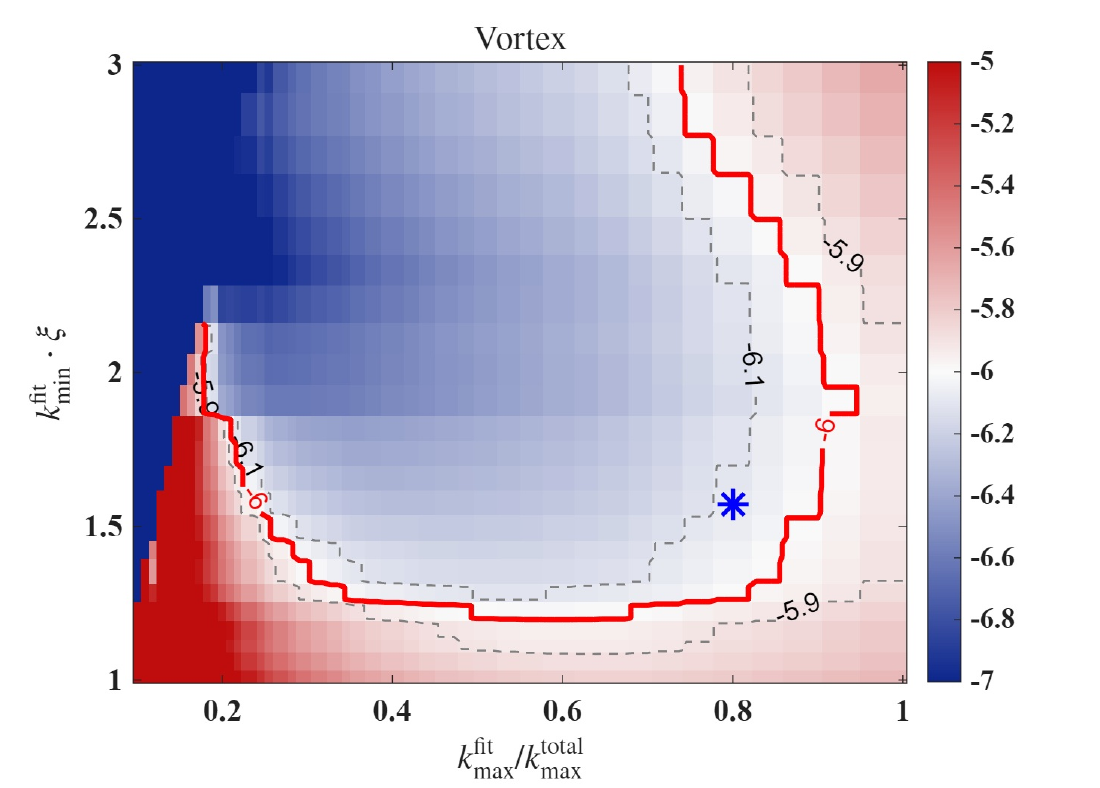}
    \includegraphics[width=0.45\columnwidth]{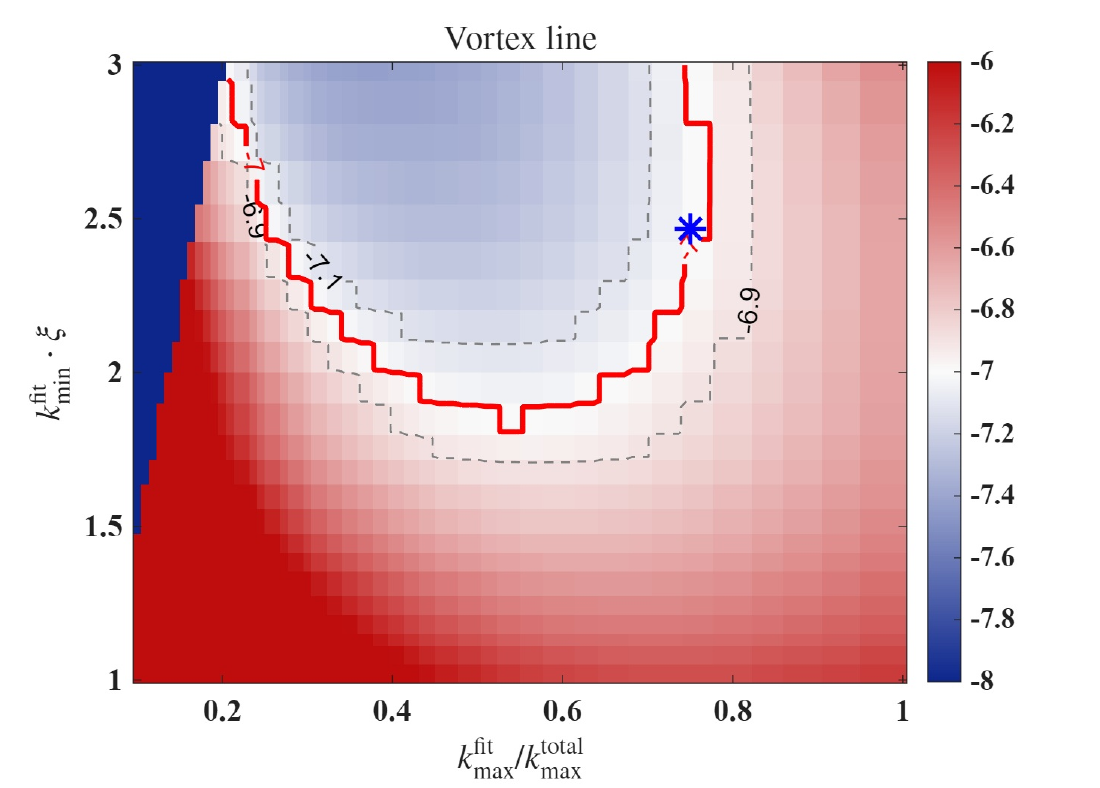}
    \caption{Analysis of the fitting interval for holographic model.
    }\label{fittingHitU1holographic}
\end{figure}

\begin{figure}
    \centering
    \includegraphics[width=0.45\columnwidth]{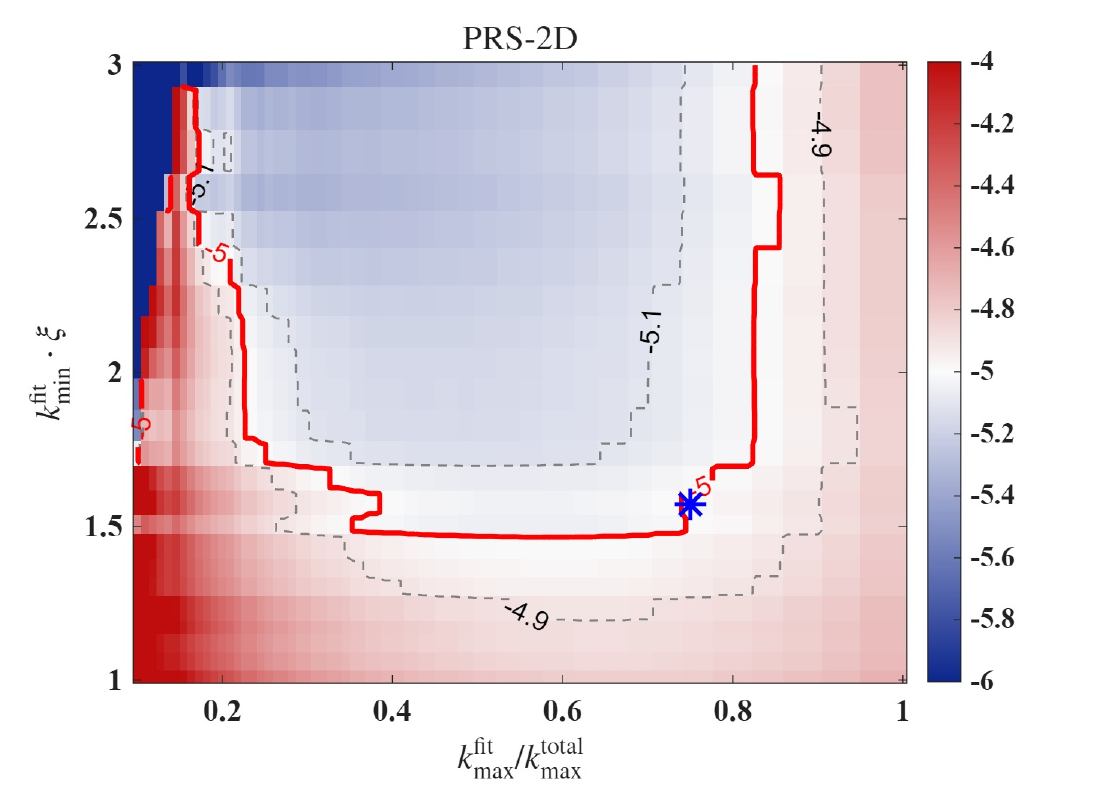}
    \includegraphics[width=0.45\columnwidth]{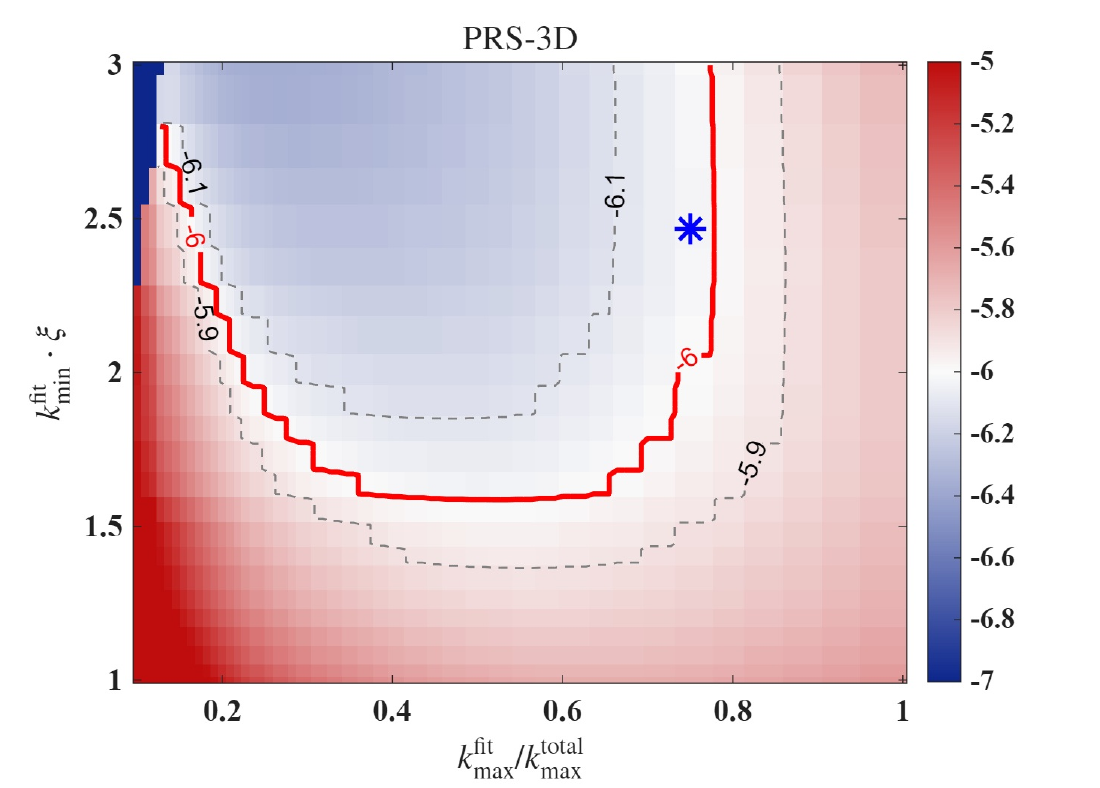}

    \caption{Analysis of the fitting interval for KG equation in the FRW background. Here we only present the fitting results under the PRS convention.
    }\label{fittingHitPRW}
\end{figure}

\end{appendix}

\end{document}